\documentclass[aps,prb,reprint,superscriptaddress]{revtex4-2}

\usepackage[T1]{fontenc}
\usepackage{graphicx}
\usepackage[tbtags]{amsmath}
\usepackage{amssymb} % For blackboard math font.

\usepackage{newtxtext,newtxmath}
\usepackage{bm}

\usepackage{xcolor}
\usepackage[hidelinks,colorlinks=true, allcolors=blue,linktoc=page]{hyperref}
\newcommand{\refsuppinfo}{the Supplemental Material \cite{SuppInfo}}

\newcommand{\reffig}{Fig.}
\newcommand{\Reffig}{Figure}
\newcommand{\reffigs}{Figs.}
\newcommand{\Reffigs}{Figures}
\newcommand{\refeq}{Eq.}

\newcommand{\refeqs}{Eqs.}

\newcommand{\eqcomma}{,}
\newcommand{\eqperiod}{.}
\newcommand{\refcite}[1]{Ref.~\cite{#1}}
\newcommand{\refeqnum}[1]{(\ref{#1})}
\newcommand{\refsec}{Sec.}

\makeatletter
\renewcommand\section{\@startsection{section}{1}{\z@}
    {0.5cm\@plus1ex\@minus0.2ex}
    {0.2cm}
    {\normalfont\small\bfseries\centering}}
\renewcommand\subsection{\@startsection{subsection}{2}{\z@}
    {0.5cm\@plus1ex\@minus0.2ex}
    {0.2cm}
    {\normalfont\small\bfseries\centering}}
\makeatother
\begin{document}

\title{Spin dependence of charge dynamics and group velocity in chiral molecules}
\thanks{This article was accepted for publication in Physical Review B on 19 May 2026. DOI: 10.1103/1g18-y6f1. Copyright \copyright\ 2026 by the American Physical Society.}

\author{Riley Stuermer}
\author{Collin VanEssen}
\author{Jacob Byers}
\author{Keith Ferrer}
\affiliation{Department of Electrical and Computer Engineering, University of Alberta, Edmonton, Alberta T6G 2V4, Canada}
\author{Prasad Gudem}
\affiliation{Department of Electrical and Computer Engineering, University of California at San Diego, La Jolla, California 92093, USA}
\author{Diego Kienle}
\affiliation{81245 Munich, Germany}
\author{Jonas Fransson}
\affiliation
{Department of Physics and Astronomy, Uppsala University, 75120 Uppsala, Sweden}
\author{Mani Vaidyanathan}
\email[Contact author: ]{maniv@ualberta.ca}
\affiliation{Department of Electrical and Computer Engineering, University of Alberta, Edmonton, Alberta T6G 2V4, Canada}

\begin{abstract}

Chiral molecules are known to preferentially select electrons with a particular spin state, an effect termed chirality-induced spin selectivity (CISS).
In this work, the transient CISS dynamics in a chiral molecule are investigated through time-dependent quantum-transport simulations, an important step toward further understanding CISS and its application in devices such as magnetoresistive random access memories and spin-based quantum computers.
We show that the spin-dependent group velocity of electrons is a possible contributor to a nonzero occupancy-based spin polarization throughout the chiral molecule.
Contrary to the case which a chiral molecule is connected to a single lead, this spin polarization persists into the steady state when two leads are connected.
We show that the simulated spin polarization qualitatively agrees with a reference experiment, as evidenced by the distinct magnetic-field signatures calculated from the spin polarization within a monolayer of chiral molecules.

\end{abstract}

\maketitle

\section{Introduction}

The ability of chiral molecules to preferentially select electrons with a certain spin state has been a topic of growing interest since its early observation \cite{1999_Ray_AsymmetricScatteringPolarizedElectrons}.
This effect, termed chirality-induced spin selectivity (CISS), has since been observed in a variety of materials, including oligopeptides \cite{2015_Kettner_SpinFilteringElectronTransport,2017_Aragones_MeasuringSpin-polarizationPower,2017_Kumar_ChiralityInducedSpinPolarizationPlacesSymmetry,2014_BenDor_LocalLightInducedMagnetizationUsingNanodots,2017_BenDor_MagnetizationSwitchingFerromagnets,2016_EckshtainLevi_ColdDenaturationInducesInversion,2022_Das_TemperatureDependentChiralInducedSpinSelectivityEffect,2020_Mishra_LengthDependentElectronSpinPolarization}, DNA \cite{2011_Gohler_SpinSelectivityElectronTransmission,2020_Mishra_LengthDependentElectronSpinPolarization,2022_Das_TemperatureDependentChiralInducedSpinSelectivityEffect}, composites such as carbon nanotubes wrapped with DNA \cite{2020_Rahman_CarrierTransportEngineeringCarbon,2015_Alam_SpinFilteringSingleWallCarbon}, and even semiconductors \cite{2016_Bloom_SpinSelectiveChargeTransport}.
The CISS effect is often quantified by the normalized difference between observables associated with the transmission of electrons with opposite spin states.
CISS literature typically denotes such values as a \textit{spin polarization}.
Common observables used for spin-polarization measurements include the spin-resolved intensity of photoelectrons transmitted through a chiral material \cite{1999_Ray_AsymmetricScatteringPolarizedElectrons,2015_Kettner_SpinFilteringElectronTransport,2011_Gohler_SpinSelectivityElectronTransmission} or the current through a chiral molecule when electrons with a particular spin state are preferentially injected by a ferromagnetic lead \cite{2020_Mishra_LengthDependentElectronSpinPolarization,2020_Rahman_CarrierTransportEngineeringCarbon,2015_Alam_SpinFilteringSingleWallCarbon,2016_Bloom_SpinSelectiveChargeTransport,2022_Das_TemperatureDependentChiralInducedSpinSelectivityEffect}.
This spin-polarization definition differs from the formal spin-polarization definition used in the remainder of this work, i.e., the normalized difference in the occupancy of electrons with opposite spin states \cite{2024_Bloom_ChiralInducedSpinSelectivity}.
The definition we adopt has been quantified in chiral molecules. 
For example, Hall sensors have been used to measure the magnetic field arising in a monolayer of chiral molecules \cite{2017_Kumar_ChiralityInducedSpinPolarizationPlacesSymmetry,2016_EckshtainLevi_ColdDenaturationInducesInversion} or the magnetization imprinted on a ferroelectric by such a monolayer \cite{2014_BenDor_LocalLightInducedMagnetizationUsingNanodots,2017_BenDor_MagnetizationSwitchingFerromagnets}, and electron paramagnetic resonance has been employed for a single chiral molecule \cite{2022_Privitera_DirectDetectionSpinPolarization}.
In contrast to existing CISS measurements associated with transmission, such approaches have captured transient data showing time-dependent spin-polarization characteristics based on occupancy \cite{2017_Kumar_ChiralityInducedSpinPolarizationPlacesSymmetry,2022_Privitera_DirectDetectionSpinPolarization}.
Importantly, we note that occupancy- and transmission-based spin-polarization constitute distinct variables constrained by the system (analogous to electron density and current);
knowing one quantity does not allow the unique determination of the other, and the most suitable definition consistent with calculated or measured observables should be chosen.

Experimental results quantifying CISS vary widely in magnitude depending on the chiral material and the measurement method.
Moreover, experiments with different molecules have shown contradictory trends with varying temperature \cite{2016_Bloom_SpinSelectiveChargeTransport,2020_Rahman_CarrierTransportEngineeringCarbon,2022_Das_TemperatureDependentChiralInducedSpinSelectivityEffect}.
To explain these observations and provide insight into the underlying mechanism of CISS, numerous theoretical models have been employed \cite{2022_Privitera_DirectDetectionSpinPolarization,2025_Smorka_InfluenceNonequilibriumVibrationalDynamics,2020_Zhang_ChiralInducedSpinSelectivityPolaron,2021_Hoff_ChiralityInducedPropagationVelocityAsymmetry,2022_Fransson_ChargeSpinDynamics,2019_Dalum_TheoryChiralInducedSpin,2018_Perlitz_HelicalLiquidCarbonNanotubes,2020_Geyer_EffectiveHamiltonianModelHelically,2019_Geyer_ChiralityInducedSpinSelectivity,2019_Fransson_ChiralityInducedSpinSelectivityRole,2024_Chiesa_ManyBodyModelsChiralityInducedSpin,2023_Huisman_ChiralityInducedSpinSelectivityCISSEffect,2020_Fransson_VibrationalOriginExchangeSplitting,2020_Ghazaryan_AnalyticModelChiralInducedSpin,2014_Guo_SpinDependentElectronTransportProtein,2021_Huisman_CISSEffectMagnetoresistance,2021_Liu_ChiralityDrivenTopologicalElectronicStructure,2019_Yang_SpinDependentElectronTransmissionModel}.
These studies propose a variety of model characteristics important to reproduce experimental CISS results, including interface effects \cite{2019_Dalum_TheoryChiralInducedSpin} and some form of inelastic scattering, such as electron-electron interactions \cite{2019_Fransson_ChiralityInducedSpinSelectivityRole,2024_Chiesa_ManyBodyModelsChiralityInducedSpin,2023_Huisman_ChiralityInducedSpinSelectivityCISSEffect} or electron-phonon interactions \cite{2020_Fransson_VibrationalOriginExchangeSplitting,2022_Fransson_ChargeSpinDynamics}. 
However, only recent theoretical studies have considered the transient behavior of CISS \cite{2024_Chiesa_ManyBodyModelsChiralityInducedSpin,2022_Fransson_ChargeSpinDynamics,2025_Smorka_InfluenceNonequilibriumVibrationalDynamics,2022_Privitera_DirectDetectionSpinPolarization,2020_Zhang_ChiralInducedSpinSelectivityPolaron,2021_Hoff_ChiralityInducedPropagationVelocityAsymmetry,2025_Zhang_DynamicalTheoryChiralInducedSpin}, leaving time-dependent CISS experiments open for further investigation \cite{2022_Evers_TheoryChiralityInducedSpin}.
A thorough understanding of CISS in the transient regime would advance the CISS effect towards applications that reduce the power requirements of magnetoresistive random access memory \cite{2017_BenDor_MagnetizationSwitchingFerromagnets} or facilitate higher-temperature qubit initialization in spin-based quantum computers \cite{2023_Chiesa_ChiralityInducedSpinSelectivityEnabling}. 
Both of these applications inherently involve transient processes.

In this study, we perform time-dependent quantum transport calculations for a chiral molecule represented by a tight-binding model connected to one or two semi-infinite leads.
To efficiently model the charge dynamics of a molecule with an atomic-site count that reflects the typical size of oligopeptides used in experiments, we use the time-dependent Landauer--B{\"u}ttiker (TDLB) formalism \cite{2014_Tuovinen_TimeDependentLandauerButtikerFormulaApplicationa,2013_Tuovinen_TimeDependentLandauer-ButtikerFormulaTransient} to describe the transient spin-polarized transport and neglect interactions.
Our simulation results demonstrate how the difference in the group velocity of electrons with different spin states leads to a nonzero spin polarization.
An asymmetric group velocity has been previously discussed for chiral molecules, but for a molecule that was effectively infinitely long because of the use of periodic boundary conditions \cite{2021_Hoff_ChiralityInducedPropagationVelocityAsymmetry} and for a molecule linked to a donor and acceptor \cite{2025_Zhang_DynamicalTheoryChiralInducedSpin}, both of which do not consider leads.
Our results corroborate the findings in these works in the presence of leads, and further demonstrate how net transport between two leads facilitates a steady-state spin polarization.
We also show a direct comparison with experiment \cite{2017_Kumar_ChiralityInducedSpinPolarizationPlacesSymmetry}.
Specifically, by calculating the magnetic field arising from the spin polarization within a monolayer of chiral molecules in the absence of interactions, we show that the group-velocity asymmetry is sufficient to qualitatively reproduce and explain the experimental time-dependent trends that \citeauthor{2017_Kumar_ChiralityInducedSpinPolarizationPlacesSymmetry} \cite{2017_Kumar_ChiralityInducedSpinPolarizationPlacesSymmetry} obtained for a monolayer of oligopeptides.
To the best of our knowledge, theory has not yet considered the magnetic field produced by electrons within a monolayer of chiral molecules, with recent progress limited to calculating the magnetic field arising from the quantum dynamics of a single chiral molecule \cite{2026_Upadhyay_ChiralityDrivenMagnetizationEmergesRelativistic}.

\section{Model}\label{Sec::Model}

\subsection{Tight-binding model of chiral molecule}

We use a simplified version of a tight-binding model \cite{2019_Fransson_ChiralityInducedSpinSelectivityRole} for a chiral molecule with $M$ helical segments of radius $a$ and pitch $l$. Each helical segment contains $N$ atomic sites, resulting in a total of $\mathbb{M}=NM$ sites.
Site $m$ has coordinates $\bm{r}_m=[a\cos\phi_m,\; a\sin\phi_m,\; l (m-1) / N]$, where $\phi_m = s 2 \pi (m-1) / N$, with $s=+1$ and $s=-1$ for right- and left-handed chiral molecules, respectively.
The corresponding Hamiltonian in second-quantized form is given by
\begin{equation}\label{Eq::MoleculeHamiltonian}
    \begin{split}
        \bm{H}_{\text{mol}} ={}&\varepsilon_0 \sum_{m=1}^{\mathbb{M}} \psi_{m}^{\dagger} \psi_{m} - t_0 \sum_{m=1}^{\mathbb{M}-1} (\psi_{m}^{\dagger} \psi_{m+1} + \text{H.c.}) \\
        &+ \lambda_0 \sum_{m=1}^{\mathbb{M}-2} (i \psi_{m}^{\dagger} \bm{\chi}_m \cdot \bm{\sigma} \psi_{m+2} + \text{H.c.})\eqperiod{}
    \end{split}
\end{equation}
Here, $\psi_m = [c_{m,+},\; c_{m,-}]^T$, where $c^\dagger_{m,\pm}$ ($c_{m,\pm}$) is the creation (annihilation) operator for the single-electron basis state $|m\rangle \otimes |\pm\rangle = |m,\pm\rangle$.
State $|m,\pm\rangle$ represents an electron at site $m$ with a spin parallel to the $\pm\hat{\bm{z}}$-axis, where $+\hat{\bm{z}}$ is the transport direction.

In \refeq{}~\refeqnum{Eq::MoleculeHamiltonian}, $\varepsilon_0$ denotes the on-site energy, and $t_0$ defines the nearest-neighbor coupling.
Spin-orbit coupling (SOC) is parameterized by $\lambda_0$ according to next-nearest-neighbor processes $i \psi_{m}^{\dagger} \bm{\chi}_m \cdot \bm{\sigma} \psi_{m+2}$, where $\bm{\chi}_m = \hat{\bm{d}}_{m+1} \times \hat{\bm{d}}_{m+2}$ quantifies the molecule's chirality, $\hat{\bm{d}}_{m+\nu} = (\bm{r}_m - \bm{r}_{m+\nu})/|\bm{r}_m - \bm{r}_{m+\nu}|$, and $\bm{\sigma} = [\sigma^x,\; \sigma^y,\; \sigma^z]$ is the vector of Pauli matrices.
Contrary to prior uses of this model \cite{2019_Fransson_ChiralityInducedSpinSelectivityRole,2020_Fransson_VibrationalOriginExchangeSplitting}, particle interactions are neglected in this work.

Semi-infinite source and drain leads are coupled to the molecule at sites 1 and $\mathbb{M}$, respectively.
These leads allow for the gain and loss of electron probability in the molecule, which is captured by adding the self-energies of the source and drain leads, $\bm{\Sigma}_{\mathrm{S}}$ and $\bm{\Sigma}_{\mathrm{D}}$, respectively, to $\bm{H}_{\text{mol}}$.
The result, termed an effective Hamiltonian, is
\begin{equation}\label{Eq::EffectiveHamiltonian}
    \bm{H}_{\mathrm{eff}} = \bm{H}_{\mathrm{mol}} + \bm{\Sigma}_{\mathrm{S}} + \bm{\Sigma}_{\mathrm{D}}\eqperiod{}
\end{equation}

We assume that the lead self-energies are energy independent, known as the \textit{wide-band-limit approximation}.
This approximation is valid when the lead density of states remains flat over the applied bias range, which is reasonable for a molecule adsorbed on gold \cite{2013_Verzijl_ApplicabilityWideBandLimit,2018_Covito_TransientChargeEnergyFlow}.
In the wide-band limit, the self-energy of lead $\alpha\in\{\mathrm{S},\,\mathrm{D}\}$ is given by $\bm{\Sigma}_{\alpha} = -i\bm{\Gamma}_{\alpha}/2$, where $\bm{\Gamma}_{\alpha}$ is an energy-independent broadening matrix that quantifies the coupling between lead $\alpha$ and each atomic site of the molecule.
The source lead is only coupled to site $1$, so $\bm{\Gamma}_{\mathrm{S}} = \Gamma_0 \psi^{\dagger}_1\psi_1$, where $\Gamma_0$ is a constant that defines the spin-independent coupling strength between the source lead and site $1$.
Similarly, for the drain lead, $\bm{\Gamma}_{\mathrm{D}} = \Gamma_0 \psi^{\dagger}_{\mathbb{M}}\psi_{\mathbb{M}}$.
A one-lead system with only the source lead connected has $\bm{\Gamma}_{\mathrm{D}}=\bm{0}$ (and hence $\bm{\Sigma}_{\mathrm{D}}=\bm{0}$).

\subsection{Time-dependent Landauer--B{\"u}ttiker formalism}

The TDLB formalism, presented by \citeauthor{2013_Tuovinen_TimeDependentLandauer-ButtikerFormulaTransient} \cite{2013_Tuovinen_TimeDependentLandauer-ButtikerFormulaTransient,2014_Tuovinen_TimeDependentLandauerButtikerFormulaApplicationa}, provides a computationally efficient method to calculate electron occupancies in tight-binding systems, facilitating the study of large molecules.

As a brief review, the TDLB formula provides an exact expression for the time-dependent one-particle reduced density matrix $n(t) = -iG^<_{\mathrm{mol}}(t,\,t)$ for a noninteracting molecule coupled to an arbitrary number of wide-band-limit leads, where lead $\alpha$ becomes biased to a potential energy $E_{\alpha} = -eV_{\alpha}$ at time $t=0$.
The TDLB formula is explicit because results at time $t$ do not rely on information from earlier times, besides the initial state at $t<0$.
At time $t$, the expected density of electrons in spin state $|\pm\rangle$ occupying site $m$ is given by
\cite{2014_Tuovinen_TimeDependentLandauerButtikerFormulaApplicationa}
\begin{equation}\label{Eq::Occupancy}
    \begin{split}
        n_{m,\pm}(t) &= \langle m,\pm| n(t) |m, \pm \rangle \\
        &= \sum_{j,k} \frac{\langle m, \pm | \Psi_j^{\mathrm{R}} \rangle}{\langle \Psi_j^{\mathrm{L}} | \Psi_j^{\mathrm{R}} \rangle} \, \frac{\langle \Psi_k^{\mathrm{R}} | m, \pm \rangle}{\langle \Psi_k^{\mathrm{R}} | \Psi_k^{\mathrm{L}} \rangle} n_{jk}(t)\eqcomma{}
    \end{split}
\end{equation}
where $\langle \Psi^{\mathrm{L}}_j |$ and $|\Psi^{\mathrm{R}}_j\rangle$ are left and right eigenvectors of $\bm{H}_{\mathrm{eff}}$, respectively, which have the same eigenvalue $\epsilon_j$ according to $\langle \Psi^{\mathrm{L}}_j|\bm{H}_{\mathrm{eff}} = \epsilon_j \langle \Psi^{\mathrm{L}}_j | $ and $\bm{H}_{\mathrm{eff}} |\Psi^{\mathrm{R}}_j\rangle = \epsilon_j |\Psi^{\mathrm{R}}_j \rangle$.
Because $\bm{H}_{\mathrm{eff}}$ is non-Hermitian, $|\Psi^{\mathrm{L}}_j \rangle$ and $|\Psi^{\mathrm{R}}_j\rangle$ are not necessarily equal.
The resulting matrix elements
\begin{equation}\label{Eq::OccupancyFactors}
    \begin{split}
        n_{jk}(t) ={}& \langle \Psi_j^{\mathrm{L}} | n(t) | \Psi_k^{\mathrm{L}} \rangle = \sum_{\alpha} \langle \Psi_j^{\mathrm{L}} | \Gamma_{\alpha} | \Psi_k^{\mathrm{L}} \rangle \\
        &\times \{ \Lambda_{\alpha,jk} + E_{\alpha}[\Pi_{\alpha,jk}(t)+\Pi^{*}_{\alpha,kj}(t)]\\
        &+ {E_{\alpha}}^2 e^{-i(\epsilon_j - \epsilon^*_k)t/\hbar} \, \Omega_{\alpha,jk}\}
    \end{split}
\end{equation}
can be written in terms of analytic expressions given by \cite{2016_Tuovinen_TimeDependentLandauer-buttikerFormalismSuperconducting,2019_Tuovinen_DistinguishingMajoranaZeroModes}
\begin{equation}\label{Eq::Lambda}
    \Lambda_{\alpha,jk} = i \frac{f(\epsilon^*_k - \mu_{\alpha})}{\epsilon^*_k-\epsilon_j} + \frac{\psi [\Phi_{\alpha}(\epsilon^*_k )] - \psi [\Phi_{\alpha}(\epsilon_j )]}{2\pi (\epsilon^*_k-\epsilon_j)}\eqcomma{}
\end{equation}
\begin{equation}\label{Eq::Pi}
    \begin{split}
        \Pi_{\alpha,jk}(t) ={}& i \frac{e^{-i(\epsilon_j - \epsilon^*_k )t/\hbar} f(\epsilon^*_k - \mu_{\alpha})}{(\epsilon^*_k-\epsilon_j)(\epsilon^*_k-\epsilon_j-E_{\alpha})} + \frac{1}{(2 \pi)^3(k_{\mathrm{B}} T)^2} \\
        &\times \sum^{\infty}_{\nu=0} \frac{e^{-2 \pi k_{\mathrm{B}} T [\nu+\Phi_{\alpha}(\epsilon_j)] t/\hbar}}{[\nu+\Phi_{\alpha}(\epsilon_j)][\nu+\Phi_{\alpha}(\epsilon^*_k)][\nu+\Phi(\epsilon_j)]}\eqcomma{}
    \end{split}
\end{equation}
and
\begin{equation}\label{Eq::Omega}
    \begin{split}
    \Omega_{\alpha,jk} ={}& \frac{1}{(\epsilon^*_k-\epsilon_j)E_{\alpha}} \\
    &\times \bigg\{i\frac{f(\epsilon^*_k-\mu)}{(\epsilon^*_k-\epsilon_j + E_{\alpha})} + \frac{\psi [\Phi(\epsilon^*_k)] - \psi [\Phi_{\alpha}(\epsilon_j)]}{2\pi (\epsilon^*_k-\epsilon_j + E_{\alpha})} \\
    & - i\frac{f(\epsilon^*_k-\mu_{\alpha})}{(\epsilon^*_k-\epsilon_j-E_{\alpha})} - \frac{\psi [\Phi_{\alpha}(\epsilon^*_k)] - \psi [\Phi(\epsilon_j)]}{2\pi (\epsilon^*_k-\epsilon_j-E_{\alpha})} \bigg\}\eqcomma{}
    \end{split}
\end{equation}
where $\mu_{\alpha} = \mu + E_{\alpha}$, $\Phi(E) = \frac{1}{2} - \frac{E - \mu}{2 \pi i k_{\mathrm{B}} T}$, $\Phi_{\alpha}(E) = \Phi(E-E_{\alpha})$, $\mu$ is the Fermi level, $f(E) = ( \exp \frac{E}{k_{\mathrm{B}} T} + 1)^{-1}$ is the Fermi function, $\psi$ is the digamma function, and $T>0$ is the electron temperature.
Typically, $\Pi_{\alpha,jk}(t)$ is defined in terms of hypergeometric functions \cite{2016_Tuovinen_TimeDependentLandauer-buttikerFormalismSuperconducting}.
In contrast, we take $\Pi_{\alpha,jk}(t)$ as a single series with a convergence rate that increases as $T$ or $t$ increase.
This result, given by \refeq{}~\refeqnum{Eq::Pi}, is an intermediate step in the analysis that leads to the established result containing hypergeometric functions \cite{2019_Tuovinen_DistinguishingMajoranaZeroModes}.
However, as detailed in \refsec{}~S1 of \refsuppinfo{}, we show that the direct use of \refeq{}~\refeqnum{Eq::Pi} can optimize the computation of $\Pi_{\alpha,jk}(t)$, which is the rate-limiting step for using the TDLB formula at $T>0$.

Because \refeq{}~\refeqnum{Eq::Occupancy} provides an expectation value, all numerical results presented in this work correspond to ensemble measurements.

\subsection{Model parameters}\label{Sec::Parameters}

The experiment conducted by \citeauthor{2017_Kumar_ChiralityInducedSpinPolarizationPlacesSymmetry} \cite{2017_Kumar_ChiralityInducedSpinPolarizationPlacesSymmetry} serves as a reference for comparison with our calculations.
Unless otherwise indicated, we use the following model parameters for our study, which were chosen based on characteristics of the relevant oligopeptides.

The oligopeptides from the reference experiment consisted of repeating Ala and 2-methylalanine (Aib) residues.
To match the dimensions of an $\alpha$ helix \cite{2008_Takeda_EffectsMonolayerStructures,1998_Fujita_MacrodipoleInteractionHelicalPeptides}, a probable geometry for an Ala-Aib chain \cite{1993_Otoda_ChainLengthDependentTransition}, we took $a=0.465\;\text{nm}$ and $l=0.540\;\text{nm}$.
The number of atomic sites per helical segment was taken as $N = 3(3.6)$, which accounts for the three backbone atoms in each residue \cite{1997_Skourtis_ElectronTransferContactMaps} and the 3.6 residues per $\alpha$-helix pitch.
The longest oligopeptides from the reference experiment consisted of seven repetitions of Ala-Aib \cite{2017_Kumar_ChiralityInducedSpinPolarizationPlacesSymmetry}, so the number of helical segments was taken as $M=2(7)/3.6$, ensuring $\mathbb{M}=42$ is an integer.
Most of our calculations used $s=+1$, which models a right-handed chiral molecule.

The Fermi level was set equal to the on-site energy, $\mu=\varepsilon_0=0\;\mathrm{eV}$, such that half of the electronic states are filled at zero bias \cite{2014_Tuovinen_TimeDependentLandauerButtikerFormulaApplicationa}.
Nearest-neighbor coupling was taken as $t_0=0.2\;\mathrm{eV}$, which is reasonable for peptides \cite{2014_Oliveira_ElectronicTransportOligopeptideChains,2006_Santhanamoorthi_ChargeTransferPolypeptidesEffect}, and the SOC parameter was taken as $\lambda_0 = 5\;\mathrm{meV}$ so that the resulting $\lambda_0/t_0$ ratio \cite{2020_Fransson_VibrationalOriginExchangeSplitting} is similar to that of a previous $\alpha$-helix model \cite{2014_Guo_SpinDependentElectronTransportProtein}.
As described in \refsec{}~\ref{Sec::Experiment}, charge transport in the reference experiment is driven electrostatically without coupling the oligopeptides to leads.
To deemphasize the influence of leads within our model, we take $\Gamma_0=0.05\;\mathrm{eV}$, which is an order of magnitude weaker than typical coupling values for gold leads \cite{2013_Verzijl_ApplicabilityWideBandLimit}.
The resulting $\Gamma_0/t_0$ ratio matches previous calculations \cite{2020_Fransson_VibrationalOriginExchangeSplitting}.
After $t=0$, the source potential energy was set to $E_{\mathrm{S}} = -eV_{\text{S}} = 0.5\;\mathrm{eV}$ so that every electronic state with energy above $\mu=\varepsilon_0$ contributes to transport under bias.
For a two-lead system, the drain potential energy was maintained at $E_{\mathrm{D}} = -eV_{\text{D}} = 0\;\mathrm{eV}$.
Finally, we used an electron temperature of $T=300\;\mathrm{K}$.

\section{Results and discussion}

\subsection{Spin polarization and group velocity\\ for molecule with one lead}\label{Sec::OneLead}

In this section, we demonstrate that asymmetry in the group velocity of electrons with different spin states facilitates a nonzero spin polarization.

\begin{figure*}[htbp!]
    \includegraphics{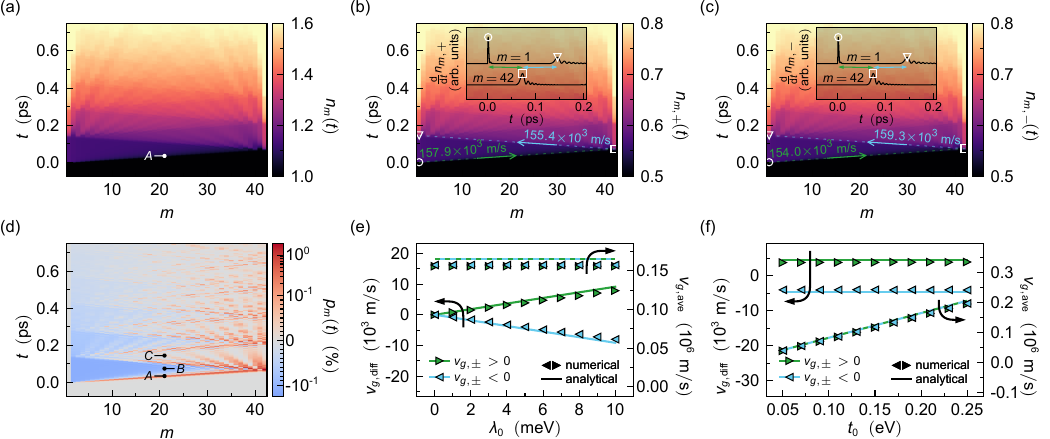}
    \caption{
        (a) Density and (d) spin polarization of electrons throughout a one-lead chiral molecule resulting from a transient calculation following application of a source potential $V_{\text{S}} = -0.5\;\text{V}$ at $t=0$.
        Markers $A$, $B$, and $C$ used in (a) and (d) are referenced by the main text.
        Electron density for (b) spin-$|+\rangle$ and (c) spin-$|-\rangle$ electrons, where group velocities are marked based on the density derivatives of site $1$ and $\mathbb{M}$ from the insets.
        Quantities in (a)--(d) are plotted as color versus time and site position.
        To improve visibility, the colormaps of $n_m(t)$ and $n_{m,\pm}(t)$ were capped at 1.6 and 0.8, respectively.
        The average group velocity magnitude of spin-$|+\rangle$ and spin-$|-\rangle$ beams, denoted $v_{g,\text{ave}}$, and the difference in group velocity magnitudes of spin-$|+\rangle$ and spin-$|-\rangle$ beams, denoted $v_{g,\text{diff}}$, are shown as a function of (e) $\lambda_0$ and (f) $t_0$, where numerical results from transient calculations are compared with analytical results.
        Throughout, quantities related to $+\hat{\bm{z}}$- and $-\hat{\bm{z}}$-moving beams are indicated by green and cyan markers, respectively.
        All unvaried parameters were set according to \refsec{}~\ref{Sec::Parameters}.
    }
    \label{Fig::OneLead}
\end{figure*}

The time-dependent spin-resolved electron density for a right-chiral molecule with only the source lead (i.e., with $\Gamma_{\mathrm{D}} = \bm{0}$) was calculated using \refeq{}~\refeqnum{Eq::Occupancy}.
Then, the density and spin polarization of electrons at site $m$ were calculated as $n_m(t) = n_{m,+}(t) + n_{m,-}(t)$ and $p_m(t) = [n_{m,+}(t) - n_{m,-}(t)]/n_m(t)$, respectively.
The time-resolved electron density is shown in \reffig{}~\ref{Fig::OneLead}(a), plotted as color versus time $t$ (vertically) and site position $m$ (horizontally), such that the time evolution at any site $m$ can be traced vertically as a function of $t$ by following the color legend.
\Reffig{}~\ref{Fig::OneLead}(a) illustrates that for $t<0$, before the source voltage is applied, the molecule is in equilibrium, with half of the electronic states filled, i.e., $n_m(t) = 1.0$ (black) for all $m$.
Although transport is quantum mechanical and involves electrons hopping from site to site, a semiclassical perspective can be used purely for visualization;
in this case, we consider a density of localized electrons at each site, produced by a beam of electrons injected from the source and propagating with a suitable group velocity.

After applying the source voltage, a $+\hat{\bm{z}}$-moving electron beam propagates through the molecule, filling states from site 1 to $\mathbb{M}$ (left to right).
The forefront of this beam at any time $t$ is determined by the site position $m$ of the positively sloped boundary between the black and dark-purple regions at the bottom of \reffig{}~\ref{Fig::OneLead}(a);
for example, marker $A$ indicates that when $t\approx0.035\;\mathrm{ps}$, the forefront is located at site $\mathbb{M}/2$.
When this $+\hat{\bm{z}}$-moving beam reaches the final site $\mathbb{M}$, it encounters an infinite potential barrier because of the lack of a drain lead and reflects as a $-\hat{\bm{z}}$-moving beam, which has a forefront that can be determined at any time $t$ using the negatively sloped boundary just above the positively sloped boundary  previously mentioned.
When the $-\hat{\bm{z}}$-moving beam returns to site 1, reflection occurs again;
however, the nonzero coupling between the source and site 1 causes a fraction of the impinging beam's electron density to be lost owing to transmission into the source lead.
This loss decreases the visibility of the subsequent $+\hat{\bm{z}}$-moving beam in \reffig{}~\ref{Fig::OneLead}(a).
Subsequent reflections at sites 1 and $\mathbb{M}$ contribute additional beams, which, along with a continual injection of electrons at the source, create an overall superposition of beams that comprises the total electron density buildup visible in \reffig{}~\ref{Fig::OneLead}(a).
Interference patterns are also visible, which arise from the wavelike nature of electrons;
this wavelike nature is not critical to the simplified semiclassical visualization being discussed here, but is an inherent part of our quantum transport calculations.
As the steady state approaches, the electron density in the molecule cannot change over time, which is discernible in \reffig{}~\ref{Fig::OneLead}(a) by a decrease in the visibility of the beam forefronts as $t$ increases, culminating in the same constant density at all sites $m$ for sufficiently large $t$.

At any stage of the reflection process, the group velocity of the electrons in a beam can be determined by dividing the path-integrated distance between sites 1 and $\mathbb{M}$ by the time it takes for the beam forefront to travel between these sites.
This approach is reasonable if the group velocity is approximately spatially independent, which is expected because the device potential, and hence the electronic band structure, is spatially uniform.
We define the time instant at which a beam forefront reaches site $m$ as the instant when the increase in the electron density at site $m$ is largest, i.e., when $\mathrm{d} n_m/ \mathrm{d} t$ peaks.
Applying this procedure to the spin-$|+\rangle$ and spin-$|-\rangle$ densities shown in \reffigs{}~\ref{Fig::OneLead}(b) and \ref{Fig::OneLead}(c), respectively, the time taken for the spin-resolved electron beams to travel between site 1 and $\mathbb{M}$ was determined, as illustrated in the inset of these plots.
Using these transport times, lines with slopes corresponding to the group velocities were superimposed on the colormap of these plots, coinciding with the visible beam forefronts.
Only the initial beam and the first reflected beam were considered for these group velocity calculations because interference becomes more significant for temporally distant reflections, distorting the peaks of $\mathrm{d} n_m/ \mathrm{d} t$.

The numerical group velocity results described above can be analyzed in terms of the analytical group velocities calculated for a chiral molecule, which, as shown in \refsec{}~S2 of \refsuppinfo{}, is given by
\begin{equation}\label{Eq::AnalyticalGroupVelocity}
    v_{g,\pm}(k) = \frac{1}{\hbar}\,\frac{\mathrm{d}}{\mathrm{d} k}  E_{\pm}(k)
    = \frac{2 t_0 d}{\hbar} \sin{k d} \mp \frac{4 \lambda_0 \chi^z d}{\hbar} \cos{2 k d}\eqcomma{}
\end{equation}
where $k$ is the wave vector,
\begin{equation}
    d = \sqrt{4 a^2 \sin^2 \frac{\pi}{N} + \frac{l^2}{N^2}}
\end{equation}
is the separation between neighboring atomic sites, and 
\begin{equation}
    \chi^z = \chi^z_m = s \frac{2a^2}{d} \, \frac{\sin\frac{2 \pi}{N} \big(1- \cos\frac{2 \pi}{N}\big)}{\sqrt{4 a^2 \sin^2\frac{2 \pi}{N} + 4 \frac{l^2}{N^2}}}
\end{equation} 
is the $\hat{\bm{z}}$ component of $\bm{\chi}_m$.
Clearly, for $N>2$, the sign of $\chi^z$ is determined by the sign of the chirality of the molecule, i.e., $\chi^z = s |\chi^z|$.
The derivation of \refeq{}~\refeqnum{Eq::AnalyticalGroupVelocity} relies on neglecting the $\hat{\bm{x}}$ and $\hat{\bm{y}}$ components of $\bm{\chi}_m$ to ensure that the energy bands for spin-$|+\rangle$ and spin-$|-\rangle$ electrons are separable.
This approximation was not made for any numerical calculation.

Each numerical group velocity was extracted from a beam forefront in \reffigs{}~\ref{Fig::OneLead}(b) and \ref{Fig::OneLead}(c), so the appropriate analytical group velocity for comparison is that from \refeq{}~\refeqnum{Eq::AnalyticalGroupVelocity} with the maximum magnitude and correct sign while being energetically available to contribute to transport.
The analytical group velocities with maximum magnitude are found in \refsec{}~S2 of \refsuppinfo{} as
\begin{equation}\label{Eq::SelectedGroupVelocity}
    v_{g,\pm} = \frac{2d}{\hbar}\begin{cases}
        +|t_0| \pm 2 \lambda_0 s |\chi^z|, & v_{g,\pm} > 0 \\
        -|t_0| \pm 2 \lambda_0 s |\chi^z|, & v_{g,\pm} < 0
    \end{cases}
\end{equation}
provided
\begin{equation}\label{Eq::GroupVelocityCriteria}
    |t_0| > |8 \lambda_0 \chi^z|\eqcomma{}
\end{equation}
which holds for the model parameters used in this section.
It is physically reasonable to enforce \refeq{}~\refeqnum{Eq::GroupVelocityCriteria}, which ensures that the nearest-neighbor coupling dominates over the next-nearest-neighbor coupling, because the larger separation between next-nearest atomic sites corresponds to a weaker coupling than that of nearest atomic sites.
In \refeq{}~\refeqnum{Eq::SelectedGroupVelocity}, $v_{g,\pm} > 0$ denotes a $+\hat{\bm{z}}$-moving electron beam, and $v_{g,\pm} < 0$ denotes a $-\hat{\bm{z}}$-moving electron beam.
The energy associated with the analytical group velocities of \refeq{}~\refeqnum{Eq::SelectedGroupVelocity} is $E_{\pm} = \varepsilon_0$, so the numerical group velocities only match \refeq{}~\refeqnum{Eq::SelectedGroupVelocity} if the lead bias satisfies $\mu \leq \varepsilon_0 \leq \mu + E_{\text{S}}$.

The average group velocity magnitudes for electrons of either spin,
\begin{equation}\label{Eq::AverageGroupVelocity}
    v_{g,\text{ave}} = \frac{|v_{g,+}| + |v_{g,-}|}{2} = \frac{2 |t_0| d}{\hbar}\eqcomma{}
\end{equation}
increases linearly with $|t_0|$ and is independent of $\lambda_0$.
In contrast, the difference in group velocity magnitudes for electrons with opposite spin states,
\begin{equation}\label{Eq::DifferenceGroupVelocity}
    v_{g,\text{diff}} = |v_{g,+}| - |v_{g,-}| = \frac{8 \lambda_0 s |\chi^z| d}{\hbar}
    \begin{cases}
        +1, & v_{g,\pm} > 0 \\
        -1, & v_{g,\pm} < 0
    \end{cases}\eqcomma{}
\end{equation}
increases linearly with $\lambda_0$ and is independent of $t_0$.
Equivalently, a group-velocity asymmetry exists for electrons with different spin states, and this asymmetry increases with the SOC strength $\lambda_0$.
The relationships from \refeqs{}~\refeqnum{Eq::AverageGroupVelocity} and \refeqnum{Eq::DifferenceGroupVelocity} motivate validating our numerical results by plotting $v_{g,\text{ave}}$ and $v_{g,\text{diff}}$ for varying $\lambda_0$ and $t_0$.
Results shown in \reffigs{}~\ref{Fig::OneLead}(e) and \ref{Fig::OneLead}(f) demonstrate the expected dependence on $\lambda_0$ and $t_0$.

\begin{figure}[htbp!]
    \includegraphics{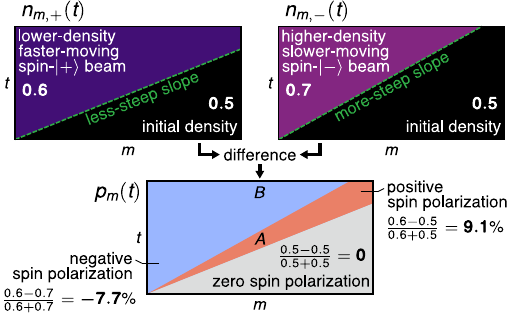}
    \caption{
        Schematic (not to scale) depicting the initial $+\hat{\bm{z}}$-moving spin-$|+\rangle$ and spin-$|-\rangle$ electron beams of \reffigs{}~\ref{Fig::OneLead}(b) and (c), respectively, and the resulting nonzero spin polarization associated with markers $A$ and $B$ of \reffig{}~\ref{Fig::OneLead}(d).
        Time and site position are on the vertical and horizontal axes, respectively.
        Density and spin-polarization values (bolded) are for illustrative purposes only.
    }
    \label{Fig::SpinPolarizationSchematic}
\end{figure}

The normalized difference between the electron densities of \reffig{}~\ref{Fig::OneLead}(b) and \reffig{}~\ref{Fig::OneLead}(c) leads to the spin-polarization plot shown in \reffig{}~\ref{Fig::OneLead}(d).
When the density of the spin-$|+\rangle$ electrons is higher (lower) than that of the spin-$|-\rangle$ electrons, the resulting spin polarization is positive (negative), which is colored as red (blue) in \reffig{}~\ref{Fig::OneLead}(d).
The spin-polarization pattern observed in \reffig{}~\ref{Fig::OneLead}(d) results from the group-velocity asymmetry given by \refeq{}~\refeqnum{Eq::DifferenceGroupVelocity} as follows: 
In a right-chiral molecule (i.e., $s=+1$), \refeq{}~\refeqnum{Eq::DifferenceGroupVelocity} indicates that the $+\hat{\bm{z}}$-moving spin-$|+\rangle$ beam injected by the source travels faster than the spin-$|-\rangle$ beam.
The spin-$|+\rangle$ beam thus has a lower electron density than the spin-$|-\rangle$ beam, because the fluxes of spin-$|+\rangle$ and spin-$|-\rangle$ electrons entering the molecule must be equal (because of the spin-independent source coupling), and the faster transport of the spin-$|+\rangle$ beam spreads these electrons more thinly.
The faster-traveling spin-$|+\rangle$ beam creates a positive spin polarization at sites not yet reached by the spin-$|-\rangle$ beam, which is discernible by the positively sloped red strip at the bottom of \reffig{}~\ref{Fig::OneLead}(d).
The instant this positive spin polarization reaches site $\mathbb{M}/2$ is indicated by marker $A$.
At a slightly later time, sites are reached by the slower-traveling spin-$|-\rangle$ beam.
Because the spin-$|-\rangle$ beam has a higher electron density than the spin-$|+\rangle$ beam, the spin polarization becomes negative.
This negative spin polarization is discernible by the blue region just above the positively sloped red strip in \reffig{}~\ref{Fig::OneLead}(d), which is indicated at site $\mathbb{M}/2$ by marker $B$.
\Reffig{}~\ref{Fig::SpinPolarizationSchematic} schematically illustrates the generation of the nonzero spin-polarization regions associated with markers $A$ and $B$ of \reffig{}~\ref{Fig::OneLead}(d).

When the $+\hat{\bm{z}}$-moving spin-$|+\rangle$ beam reaches site $\mathbb{M}$, it reflects as a $-\hat{\bm{z}}$-moving spin-$|+\rangle$ beam.
Per \refeq{}~\refeqnum{Eq::DifferenceGroupVelocity}, the $-\hat{\bm{z}}$-moving spin-$|+\rangle$ beam travels slower than the $+\hat{\bm{z}}$-moving spin-$|+\rangle$ beam.
However, the magnitude of electron flux must be conserved, so the decrease in electron speed is accompanied by a proportional increase in electron density.
Thus, the superposition of this pair of $+\hat{\bm{z}}$- and $-\hat{\bm{z}}$-moving spin-$|+\rangle$ beams has no net flux, which is equivalent to a stationary density of spin-$|+\rangle$ electrons.
Similarly, a stationary density follows for the spin-$|-\rangle$ electrons.
With no flux lost, the stationary spin-$|+\rangle$ and spin-$|-\rangle$ electrons have equal density,
which follows from symmetries in the group velocities and is illustrated by a schematic of the reflection process shown in \reffig{}~S3 in \refsec{}~S3 of \refsuppinfo{}.
Thus, the nonzero spin polarization is removed as the $-\hat{\bm{z}}$-moving spin beams return to site 1
(except for higher-order interference effects).
The removal of this spin polarization is discernible by the gray region just above the
previously mentioned blue region 
(containing marker $B$) of \reffig{}~\ref{Fig::OneLead}(d),
which is indicated at site $\mathbb{M}/2$ by marker $C$.
After the $-\hat{\bm{z}}$-moving spin beams reflect at site 1, the processes represented by markers $A$, $B$, and $C$ repeat.
However, at site 1, in addition to reflection, electrons are lost through transmission into the source lead. 
Hence, although the processes represented by markers $A$, $B$, and $C$ do repeat, the decreased electron density of the beams involved in the processes and the increased total electron density in the molecule decreases the magnitude of the spin polarization as $t$ increases, culminating in a zero spin polarization at all sites $m$ in the steady state.

\subsection{Spin polarization for molecule with two leads}\label{Sec::TwoLeads}

\begin{figure*}[htbp!]
    \includegraphics{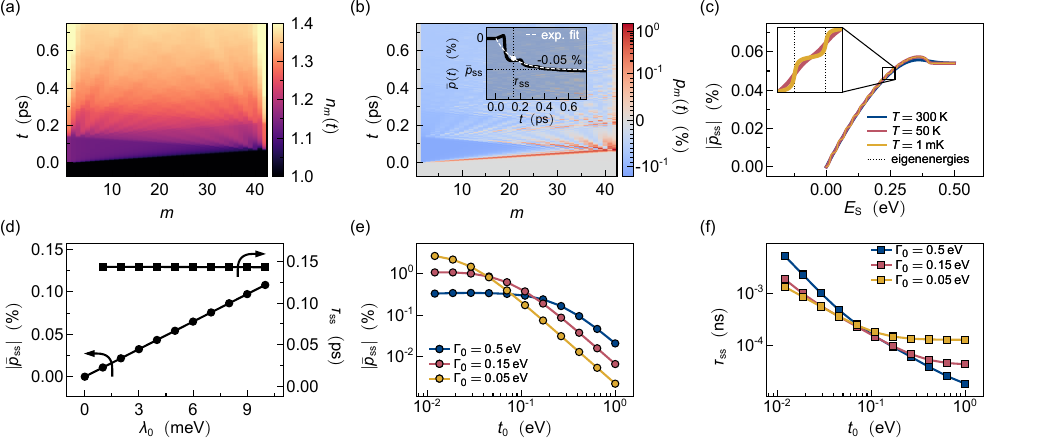}
    \caption{
        (a) Density and (b) spin polarization of electrons throughout a two-lead chiral molecule resulting from a transient calculation following application of a source potential $V_{\text{S}} = -0.5\;\text{V}$ at $t=0$.
        To improve visibility, the colormap of $n_m(t)$ was capped at 1.4.
        Inset in (b) shows the steady-state time constant $\tau_{\text{ss}}$ resulting from an exponential fit of the site-averaged spin polarization $\bar{p}(t)$, and the steady-state site-averaged spin polarization $\bar{p}_{\text{ss}}$.
        Characterization of $\bar{p}_{\text{ss}}$ as a function of (c) $E_{\text{S}}$ and $T$, (d) $\lambda_0$, and (e) $t_0$ and $\Gamma_0$.
        Characterization of $\tau_{\text{ss}}$ as a function of (d) $\lambda_0$ and (f) $t_0$ and $\Gamma_0$.
        All unvaried parameters were set according to \refsec{}~\ref{Sec::Parameters}, except for (e) and (f), where $E_{\text{S}}$ was set sufficiently high to fill all conducting states.
    }
    \label{Fig::TwoLeads}
\end{figure*}

\Reffigs{}~\ref{Fig::TwoLeads}(a) and \ref{Fig::TwoLeads}(b) show the density and spin polarization of electrons, respectively, when a right-chiral molecule is connected to both the source and drain leads.
The propagation of the electron beams is similar to the one-lead case, except that now an additional fraction of the electron density in each $+\hat{\bm{z}}$-moving beam is lost because of transmission into the drain lead.
Unlike the one-lead case, this loss prevents perfect flux cancellation between each pair of $+\hat{\bm{z}}$-moving and $-\hat{\bm{z}}$ moving beams, incident upon and reflected from the drain lead;
a residual $+\hat{\bm{z}}$-moving flux from the superposition of such beams must thus exist, i.e., from the superposition of incident and reflected beams for spin-$|+\rangle$ electrons, and the incident and reflected beams for spin-$|-\rangle$ electrons.
Furthermore, because the spin-independent source lead injects an equal flux of spin-$|+\rangle$ and spin-$|-\rangle$ electrons, and there is no spin-flip mechanism to break the symmetry of spin-resolved electron quantities \cite{2005_Kiselev_ProhibitionEquilibriumSpinCurrents}, the residual $+\hat{\bm{z}}$-moving fluxes associated with the two spins must be equal, as verified in \reffig{}~S4 in \refsec{}~S4 of \refsuppinfo{}.
Given that $+\hat{\bm{z}}$-moving spin-$|+\rangle$ electrons have a higher group velocity than $+\hat{\bm{z}}$-moving spin-$|-\rangle$ electrons, it then follows that the residual electron density must be lower for spin-$|+\rangle$ electrons than for spin-$|-\rangle$ electrons.
With sufficient time, this density difference results in a negative spin polarization being established and persisting into the steady state at all sites $m$, as seen in \reffig{}~\ref{Fig::TwoLeads}(b) and illustrated by a schematic of the reflection process shown in \reffig{}~S3 in \refsec{}~S3 of \refsuppinfo{}.

Importantly, the explanation for the spin polarization in this section and \refsec{}~\ref{Sec::OneLead} demonstrates that, ultimately, the presence of net transport, whether in the steady state or transient regime and with one or two leads, 
facilitates a nonzero spin polarization (based on occupancy, as we have defined it) even without interactions. 
However, we add that the presence of an occupancy-based spin polarization does not necessitate a transmission-based spin polarization;
in our results, the presence of spin-independent leads and the lack of a spin-flip mechanism (i.e., no interactions) guarantees \cite{2005_Kiselev_ProhibitionEquilibriumSpinCurrents} that no transmission-based spin polarization exists, despite the existence of an occupancy-based spin polarization.

For our two-lead calculation, the site-averaged spin polarization $\bar{p}(t) = \sum_{m} p_m(t) / \mathbb{M}$ starts at zero and attains a value $\bar{p}_{\mathrm{ss}}$ in the steady state, as shown in the inset of \reffig{}~\ref{Fig::TwoLeads}(b).
We quantify the spin polarization of a molecule by its steady-state value and the associated time constant to reach steady state $\tau_{\mathrm{ss}}$.
Reflections of the electron beams result in a nonmonotonic behavior of the $\bar{p}(t)$ plot, so we define $\tau_{\mathrm{ss}}$ as a parameter for an exponential fit of $\bar{p}(t)$.
By studying how $p_{\mathrm{ss}}$ and $\tau_{\mathrm{ss}}$ are affected by various model parameters, trends that maximize the spin polarization can be identified.

As seen in \reffig{}~\ref{Fig::TwoLeads}(c), a nontrivial lead bias exists that maximizes $|p_{\mathrm{ss}}|$. 
That is, $|p_{\mathrm{ss}}|$ does not increase monotonically with increasing $E_{\mathrm{S}}$.
Instead, increasing $E_{\text{S}}$ beyond $\approx0.36\;\mathrm{eV}$ slightly decreases $|p_{\mathrm{ss}}|$, which is less apparent at higher temperatures because of the thermal broadening of the Fermi function.
At low temperatures, steps in $|p_{\text{ss}}|$ appear at each eigenenergy, as seen in the enlarged inset of \reffig{}~\ref{Fig::TwoLeads}(c).
These steps illustrate that each electronic state contributing to transport impacts the spin polarization.

As seen in \reffig{}~\ref{Fig::TwoLeads}(d), $|p_{\text{ss}}|$ varies linearly with $\lambda_0$ for $0 < \lambda_0 \ll t_0,\Gamma_0$. 
This trend, along with \refeq{}~\refeqnum{Eq::DifferenceGroupVelocity} (which indicates that $v_{g,\text{diff}}$ also varies linearly with $\lambda_0$), supports the notion that a larger asymmetry in the spin-dependent group velocity results in a larger spin polarization.
\Reffig{}~\ref{Fig::TwoLeads}(d) also shows that the time for the spin polarization to reach a steady state is unaffected by $\lambda_0$ for the range of values considered because $\tau_{\text{ss}}$ remains constant.

The impact of $t_0$ and $\Gamma_0$ on the steady-state spin polarization is shown in \reffigs{}~\ref{Fig::TwoLeads}(e) and \ref{Fig::TwoLeads}(f), which quantify $|p_{\text{ss}}|$ and $\tau_{\text{ss}}$, respectively.
\Reffig{}~\ref{Fig::TwoLeads}(f) shows that $\tau_{\mathrm{ss}}$ decreases with increasing $t_0$ until $t_0$ approximately exceeds $\Gamma_0$, at which point $\tau_{\mathrm{ss}}$ remains constant.
This trend indicates that the lead coupling can be the rate-limiting step for the spin polarization to reach a steady state.
In contrast, \reffig{}~\ref{Fig::TwoLeads}(e) shows that $|p_{\text{ss}}|$ remains constant with increasing $t_0$ until $t_0$ approximately exceeds $\Gamma_0$, at which point $|p_{\text{ss}}|$ decreases with increasing $t_0$.
This trend indicates that $|p_{\mathrm{ss}}|$ is limited by either the lead coupling or neighbor coupling, whichever is stronger.

Together, the trends observed in \reffigs{}~\ref{Fig::TwoLeads}(d) and \ref{Fig::TwoLeads}(e) indicate that enhancing the steady-state spin polarization requires maximizing the SOC strength relative to other couplings, assuming that $\lambda_0$ remains sufficiently small to satisfy \refeq{}~\refeqnum{Eq::GroupVelocityCriteria}.

\subsection{Comparison to experimental time-dependent results}\label{Sec::Experiment}

\begin{figure*}[htbp!]
    \includegraphics{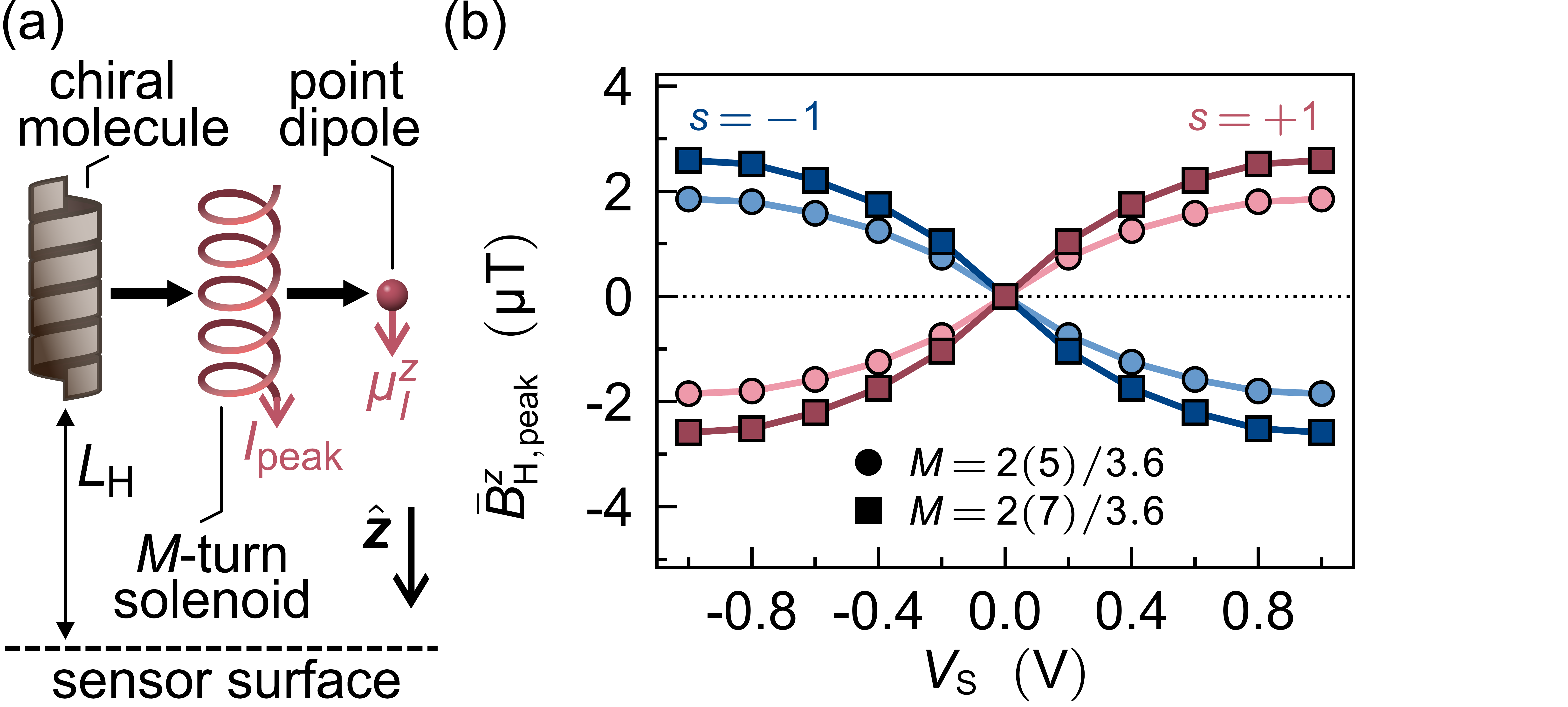}
    \caption{
        (a) Diagram (not to scale) of the reference experiment \cite{2017_Kumar_ChiralityInducedSpinPolarizationPlacesSymmetry} showing the magnetic field generated over a Hall sensor by a monolayer of chiral molecules when $V_{\text{G}}$ is applied.
        (b) Hall voltage for a monolayer of $X$-$($\textsc{l}-Ala-\textsc{l}-Aib$)_5$-$Y$ oligopeptides with $V_{\text{G}} = -10\;\text{V}$ applied at $t=0$, extracted from \refcite{2017_Kumar_ChiralityInducedSpinPolarizationPlacesSymmetry}.
        (c) Peak Hall voltage for a variety of monolayers at various $V_{\text{G}}$, extracted from \refcite{2017_Kumar_ChiralityInducedSpinPolarizationPlacesSymmetry}.
        Here, $X$-$($Ala-Aib$)_n$-$Y$ has $2n$ residues of chirality \textsc{l} or \textsc{d}, where $X=\text{SHCH\textsubscript{2}CH\textsubscript{2}CO}$ and $Y = \text{COOH}$.
        (d) Diagram (not to scale) of our monolayer model showing the magnetic field generated by chiral molecules when $V_{\text{S}}$ is applied at $t=0$.
        (e) Magnetic field from the spin polarization calculated for our monolayer model with $M=2(7)/3.6$, $s=-1$, and $V_{\text{S}}=-0.5\;\text{V}$.
        The inset shows the result after applying a low-pass filter.
        (f) Peak magnetic field from the spin polarization calculated for our monolayer model at various $M$, $s$, and $V_{\text{S}}$.
        Here, $M=2n/3.6$ accounts for $2n$ residues.
        Additional parameters are given in \refsec{}~\ref{Sec::Parameters}.
    }
    \label{Fig::Experiment}
\end{figure*}

We validate our findings from the last two sections by comparing results obtained with our model to existing experimental results.
The reference experiment conducted by \citeauthor{2017_Kumar_ChiralityInducedSpinPolarizationPlacesSymmetry} \cite{2017_Kumar_ChiralityInducedSpinPolarizationPlacesSymmetry} studied a monolayer of chiral oligopeptides adsorbed on intrinsic GaN, which is a semiconductor that is electrically insulating at room temperature in the absence of defects \cite{2020_Roccaforte_IntroductionGalliumNitrideProperties}, with side lengths $L_x = 500\;\mathrm{\upmu m}$ and $L_y = 40\;\mathrm{\upmu m}$.
Applying a voltage $V_{\mathrm{G}}$ to a gate parallel to the monolayer generates an electric field $\bm{E}_{\mathrm{G}} = E_{\mathrm{G}}\hat{\bm{z}}$ through the monolayer, where $\hat{\bm{z}}$ is collinear with the longitudinal axis of the chiral molecules.
A diagram of this experimental setup is shown in \reffig{}~\ref{Fig::Experiment}(a).
Initially, applying $\bm{E}_{\mathrm{G}}$ causes charge reorganization within the molecules of the monolayer, which is accompanied by a magnetic-field signature associated with a nonzero spin polarization throughout the monolayer.
After charge equilibration, transport stops, the electronic spin states randomize, and the magnetic-field signature decays.
Removing $\bm{E}_{\mathrm{G}}$ allows charge to flow again through the chiral molecule, now in the opposite direction, leading to another magnetic-field signature of opposite polarity.

In the reference experiment, the magnetic-field signatures were measured by a Hall sensor that had dimensions identical to the monolayer but offset in the $\hat{\bm{z}}$ direction by $L_{\mathrm{H}}=22\;\mathrm{nm}$ \cite{2017_Kumar_ChiralityInducedSpinPolarizationPlacesSymmetry,2020_Roccaforte_IntroductionGalliumNitrideProperties}.
The $\hat{\bm{z}}$ component of this magnetic field on the sensor surface, denoted $B^z_{\mathrm{H}}(x,\,y,\,t)$, induces a Lorentz force on an $\hat{\bm{x}}$-directed current $I_{\mathrm{H}}$ flowing through the Hall sensor, resulting in a Hall voltage $V_{\mathrm{H}}(t)$ across the $y$ dimension of the Hall sensor.
In general, the Hall voltage depends on a \textit{weighted} spatial average of the magnetic field \cite{1994_Mantorov_InterpretationHallEffectMeasurements}.
However, as justified in \refsec{}~S7 of \refsuppinfo{}, we assume $V_{\mathrm{H}}(t) \propto \bar{B}^z_{\mathrm{H}}(t)$, where $\bar{B}^z_{\mathrm{H}}(t)$ is the \textit{simple} spatial average of $B^z_{\mathrm{H}}(x,\,y,\,t)$.
A large capacitance affects the experimental measurement \cite{2017_Kumar_ChiralityInducedSpinPolarizationPlacesSymmetry}, so $V_{\mathrm{H}}(t)$ is modified by an $RC$ low-pass filter with a resistance $R$ and capacitance $C$.
We designate the resulting low-passed voltage measured by the reference experiment as $\widetilde{V}_{\mathrm{H}}(t)$.

The measured $\widetilde{V}_{\mathrm{H}}(t)$ includes the spin polarization of the monolayer as follows:
Based on the magnetic dipole moment of an electron \cite{1957_Sommerfield_MagneticDipoleMoment}, the total magnetic dipole moment of spin-$|+\rangle$ and spin-$|-\rangle$ electrons at atomic site $m$ can be written as
\begin{equation}\label{Eq::DipoleMagneticMoment}
    \begin{split}
        \bm{\mu}_m(t) &= -\mu_{\mathrm{B}} n_{m,+}(t)\hat{\bm{z}} + \mu_{\mathrm{B}} n_{m,-}(t)\hat{\bm{z}} \\
        &= -\mu_{\mathrm{B}} p_m(t) \, n_m(t) \hat{\bm{z}} = \mu_m^z(t)\hat{\bm{z}}\eqcomma{}
    \end{split}
\end{equation}
where $\mu_{\mathrm{B}}$ is the Bohr magneton.
A dipole at position $\bm{r}' = [x',\; y',\; z']$ with a magnetic moment given by \refeq{}~\refeqnum{Eq::DipoleMagneticMoment} generates a magnetic field at position $\bm{r} = [x,\,y,\,z]$ with a $\hat{\bm{z}}$ component given by \cite{1998_Heras_ExplicitExpressionsElectric}
\begin{equation}\label{Eq::DipoleMagneticField}
    \begin{split}
        B_m^z(\bm{r},\,\bm{r}',\,t) ={}& \frac{\mu_0 \mu_m^z(t)}{4\pi} \bigg[ \frac{3 (z-z')^2}{|\bm{r}-\bm{r}'|^5} - \frac{1}{|\bm{r}-\bm{r}'|^3}\bigg] \\
        &+ [\text{intermediate-zone term}]\\
        &+ [\text{far-zone term}]\eqperiod{}
    \end{split}
\end{equation}
In \refeq{}~\refeqnum{Eq::DipoleMagneticField}, we have not explicitly stated the intermediate- and far-zone terms involving time derivatives of $\mu_m^z(t)$ that appear in the exact magnetic field for a time-dependent magnetic dipole \cite{1998_Heras_ExplicitExpressionsElectric}.
Then, $B^z_{\mathrm{H}}(x,\,y,\,t)$ from the spin polarization results from summing \refeq{}~\refeqnum{Eq::DipoleMagneticField} for each site in the monolayer.
In \refeq{}~\refeqnum{Eq::DipoleMagneticField}, we have neglected any delays arising from the propagation times of magnetic fields.
This approximation is reasonable assuming that any distortion in $B^z_{\mathrm{H}}(x,\,y,\,t)$ owing to delays from spatially separated dipoles is ultimately concealed by the $RC$ low-pass filter, which itself delays  and smooths the measured signal.

The reference experiment has two key characteristics:
(1) an external force facilitates electron transport through the chiral molecules, and
(2) transport ceases with time. The one-lead model from \refsec{}~\ref{Sec::OneLead} exhibits both of these traits because the source voltage facilitates electron transport, which stops when all available states are filled.
Hence, the magnetic field resulting from the electron density calculated for many chiral molecules, all connected to one lead, provides a point of comparison to experimental results.
Two plots, adapted from experimental data \cite{2017_Kumar_ChiralityInducedSpinPolarizationPlacesSymmetry}, will serve as reference for this comparison.
The first plot, shown in \reffig{}~\ref{Fig::Experiment}(b), shows $\widetilde{V}_{\mathrm{H}}(t)$ after $V_{\mathrm{G}}$ was applied to initiate transport, where $\widetilde{V}_{\mathrm{H,peak}}$ indicates the measured peak voltage.
The second plot, shown in \reffig{}~\ref{Fig::Experiment}(c), shows how $\widetilde{V}_{\mathrm{H,peak}}$ varies with $V_{\mathrm{G}}$ for three distinct monolayers, differing in the length and chirality of their constituent molecules.

A transport calculation that includes every molecule in the monolayer would be infeasible.
Thus, using the TDLB formalism, we calculate the electron density for one molecule and use the result as an estimate for the electron density of each molecule in the monolayer.
This procedure implicitly neglects any interactions between molecules e.g., screening, but is otherwise reasonable, as shown in \refsec{}~S5 of \refsuppinfo{}.
Because the monolayer density in the reference experiment is not specified, we assume that the molecules have a hexagonal packing with an average surface density $\rho = 8.5\times10^{13}\;\mathrm{cm}^{-2}$, consistent with measurements for similar Ala-Aib oligopeptides \cite{2008_Takeda_EffectsMonolayerStructures}.
Because the distance between neighboring molecules is much smaller than the dimensions of the monolayer, we can treat the monolayer as a continuous distribution of molecules with surface density $\rho$.
This approximation is necessary because summing the time-dependent, spatially resolved magnetic-field contributions for every electron in the monolayer is numerically infeasible.
Each molecule contains $\mathbb{M}$ atoms, making the monolayer equivalent to $\mathbb{M}$ sublayers, each with a continuous atomic surface density $\rho$.
Sublayer $m$ has a uniform magnetic dipole surface density $\rho \mu^z_m(t)$ because the same transport calculation was employed for each molecule.
A diagram of our monolayer model is shown in \reffig{}~\ref{Fig::Experiment}(d).

The calculation of $\bar{B}^z_{\mathrm{H}}(t)$ for our monolayer model has been simplified to finding the $\hat{\bm{z}}$ component of the spatially averaged magnetic field arising from a uniform surface density of magnetic dipoles and summing the contributions from all $\mathbb{M}$ sublayers.
Starting with \refeq{}~\refeqnum{Eq::DipoleMagneticField}, the magnetic field arising from the surface density of dipoles in sublayer $m$ is obtained by replacing $\mu_m^z(t)$ with $\rho \mu_m^z(t) \, \mathrm{d}x' \, \mathrm{d}y'$ and integrating over the dipole surface defined by $x' \in [0, \, L_x]$, $y' \in [0, \, L_y]$, and $z' = z_m$.
An analytical approximation of the resulting integral, which is valid on a sensor surface defined by $x \in [0, \, L_x]$, $y \in [0, \, L_y]$, and $z = L_{\mathrm{H}} + l(\mathbb{M} - 1)/N$, is
\begin{equation}\label{Eq::DipoleSurface}
    \begin{split}
        B^z_{\mathrm{H},m}(x,\,y,\,t) ={}& \frac{\mu_0 \rho \mu^z_m(t)}{2 \pi} 
        \bigg[F_m(L_y-y) + F_m(y)\bigg. \\
        &+ \bigg. \frac{F_m(L_x-x)}{2} + \frac{F_m(x)}{2}\bigg]\eqcomma{}
    \end{split}
\end{equation}
where $F_m(\alpha) = \alpha / (\alpha^2 + {L_m}^2)$ with $L_m = L_{\mathrm{H}} + l(\mathbb{M} - m)/N$ specifying the $z$ separation between sublayer $m$ and the sensor surface.
Integrating \refeq{}~\refeqnum{Eq::DipoleSurface} over the sensor surface and dividing by $L_x L_y$ yields the spatially averaged value of the magnetic field from sublayer $m$, which is given by
\begin{equation}\label{Eq::DipoleSurface_Average}
    \begin{split}
        \bar{B}^z_{\mathrm{H},m}(t) ={}& \frac{\mu_0 \rho \mu^z_m(t)}{2 \pi} \bigg[\frac{1}{L_y}\ln{\bigg(\frac{{L_y}^2}{{L_m}^2} + 1\bigg)}\bigg. \\
        &+ \bigg. \frac{1}{2L_x}\ln{\bigg(\frac{{L_x}^2}{{L_m}^2} + 1\bigg)}\bigg]\eqperiod{}
    \end{split}
\end{equation}
Finally, $\bar{B}^z_{\mathrm{H}}(t) = \sum_{m=1}^{\mathbb{M}} \bar{B}^z_{\mathrm{H},m}(t)$.
The derivation of \refeq{}~\refeqnum{Eq::DipoleSurface}, which is valid provided $L_m \ll L_y < L_x$ and $\rho L_yL_x \gg 1$, is detailed in \refsec{}~S6 of \refsuppinfo{}, where we also estimate that, at our device dimensions, \refeq{}~\refeqnum{Eq::DipoleSurface_Average} gives a result within $\approx$ 5\% of the numerical result for a discrete grid of dipoles.
In deriving \refeqs{}~\refeqnum{Eq::DipoleSurface} and \ref{Eq::DipoleSurface_Average}, we neglected the intermediate- and far-zone contributions present in \refeq{}~\refeqnum{Eq::DipoleMagneticField}.
This \textit{quasistatic approximation} is shown to be reasonable for our application in \refsec{}~S8 of \refsuppinfo{}.

In general, an electron's magnetic dipole moment $\bm{\mu}_m(t) = \mu_m^x(t)\hat{\bm{x}} + \mu_m^y(t)\hat{\bm{y}} + \mu_m^z(t)\hat{\bm{z}}$ has  $\hat{\bm{x}}$ and $\hat{\bm{y}}$ components, contrary to the purely $\hat{\bm{z}}$-component spin-$|\pm\rangle$ electrons considered by \refeq{}~\refeqnum{Eq::DipoleMagneticMoment}.
However, assuming that all molecules in the monolayer are oriented with the same azimuthal angle, $\bm{\mu}_m(t)$ will have the same direction for all dipoles within a sublayer.
Then, as shown in \refsec{}~S6 of \refsuppinfo{}, $\bar{B}^z_{\text{H},m}(t)$ from $\mu_m^x(t)$ and $\mu_m^y(t)$ over sublayer $m$ is zero.
Alternatively, assuming that each molecule has a random azimuthal rotation, the spatial average of $\mu_m^x(t)$ and $\mu_m^y(t)$ over each sublayer is zero.
In either case, $\mu_m^x(t)$ and $\mu_m^y(t)$ can be neglected. 

A plot of $\bar{B}^z_{\mathrm{H}}(t)$ calculated for our monolayer model, based on the transport calculation of a one-lead chiral molecule, is shown in \reffig{}~\ref{Fig::Experiment}(e).
To emulate filtering in the experiment, we applied a low-pass filter to the calculated $\bar{B}^z_{\mathrm{H}}(t)$ signal.
The result, shown in the inset of \reffig{}~\ref{Fig::Experiment}(e), qualitatively matches the shape of the experimental result in \reffig{}~\ref{Fig::Experiment}(b).
Although the presence of Coulomb interactions and a HOMO-LUMO gap \cite{2024_Day_ChiralityInducedSpinSelectivityAnalysis} in a physical molecule may limit the amount of charge gained at long times compared to that of a calculation with our model molecule, importantly, the \textit{peak} magnetic field shown in \reffig{}~\ref{Fig::Experiment}(e) occurs shortly after transport begins, when only $4.65$ electrons ($0.11$ electrons per atomic site) have been added to the molecule.
Therefore, our reported trends based on peak magnetic field are not impacted by the excess charge gained at long times.
Similarly, we also note that, while our \reffigs{}~\ref{Fig::OneLead} and \ref{Fig::TwoLeads} include quantities at long times, we find that the qualitative trends of these figures remain unchanged by limiting the bias voltage to limit the charge, as might occur with a HOMO-LUMO gap or Coulomb interactions.

\Reffig{}~\ref{Fig::Experiment}(f) shows the peak of $\bar{B}^z_{\mathrm{H}}(t)$, denoted
$\bar{B}^z_{\mathrm{H,peak}}$, calculated for various monolayers at different $V_{\mathrm{S}}$.
These monolayers differ in the length and chirality of their constituent molecules.
This plot exhibits qualitative agreement with the trends observed in \reffig{}~\ref{Fig::Experiment}(c):
(1) the signs of $\bar{B}^z_{\mathrm{H,peak}}$ and $\widetilde{V}_{\mathrm{H,peak}}$ flip with the handedness of the chirality,
(2) the magnitudes of $\bar{B}^z_{\mathrm{H,peak}}$ and $\widetilde{V}_{\mathrm{H,peak}}$ increase with increasing molecular length,
(3) the signs of $\bar{B}^z_{\mathrm{H,peak}}$ and $\widetilde{V}_{\mathrm{H,peak}}$ flip with the sign of the external bias that facilitates electron transport, and
(4) the magnitudes of $\bar{B}^z_{\mathrm{H,peak}}$ and $\widetilde{V}_{\mathrm{H,peak}}$ increase and saturate as the magnitude of the external bias that facilitates electron transport increases.
The slight asymmetry in \reffig{}~\ref{Fig::Experiment}(f) with the sign of the external bias arises from the $\hat{\bm{x}}$ and $\hat{\bm{y}}$ components of $\bm{\chi}_m$.
This asymmetry can be reduced by increasing the radius or decreasing the pitch of our model molecule, which reduces the magnitude of the $\hat{\bm{x}}$ and $\hat{\bm{y}}$ components of $\bm{\chi}_m$ relative to that of the $\hat{\bm{z}}$ component.
Asymmetry in the experimental results of \reffig{}~\ref{Fig::Experiment}(c) could also be embedded within the error bars showing measurement uncertainty.
In both cases, any small asymmetry is of secondary importance compared to the more dominant magnetic-field trends present in both our model and experiments.

Despite the qualitative agreement, a quantitative discrepancy exists between the nanotesla magnetic-field peak shown in \reffig{}~\ref{Fig::Experiment}(e), occurring on a picosecond timescale, and the millitesla magnetic-field peaks anticipated by the reference experiment, detected on a timescale of seconds \cite{2017_Kumar_ChiralityInducedSpinPolarizationPlacesSymmetry}.
Deviation could be expected from our use of a phenomenological tight-binding model \cite{2019_Fransson_ChiralityInducedSpinSelectivityRole} instead of a first-principles model \cite{2024_Day_ChiralityInducedSpinSelectivityAnalysis}.
In addition, the experimental timescale is expected to increase because of an exchange interaction between molecules \cite{2017_Kumar_ChiralityInducedSpinPolarizationPlacesSymmetry}, which was not included in our model.
The significance of interactions between molecules is supported by the association of tighter monolayer packing with reduced electron transfer speeds \cite{2008_Takeda_EffectsMonolayerStructures}.
Because inelastic scattering has been shown to enhance spin-polarization effects \cite{2023_Huisman_ChiralityInducedSpinSelectivityCISSEffect}, our calculated magnetic-field magnitudes may be reduced by the lack of interactions within molecules in our model.
The $RC$ filter, which spreads the measured signal over time, adds complexity in making a comparison.

To contextualize the quantitative discrepancy, we use \refeq{}~\refeqnum{Eq::DipoleSurface_Average} to calculate the physical-maximum magnetic field that could be generated by a monolayer by supposing each atomic site has 100\% spin polarization.
No inclusion of interactions can increase the spin polarization in the monolayer beyond 100\%;
the corresponding result for the magnetic field is impacted only by the monolayer density and number of atomic sites.
Despite accurately estimating the monolayer density using measurements from similar monolayers \cite{2008_Takeda_EffectsMonolayerStructures}, the resulting value of $26.0\;\mathrm{\upmu T}$ for the magnetic field from this calculation (assuming one atomic site per backbone atom) remains orders of magnitude below the ${\sim}5\;\mathrm{mT}$ magnetic-field peak size anticipated by the reference experiment \cite{2017_Kumar_ChiralityInducedSpinPolarizationPlacesSymmetry}.
Note that, in this context, an increase to 192 atomic sites per backbone atom would be required to reach the experimentally anticipated  magnetic-field size, which corresponds to a clearly nonphysical $42(192)=8064$ electrons per molecule.
This suggests that an entirely different mechanism than the ordinary Hall effect is likely responsible for converting the spin polarization to the sizable Hall voltage in the reference experiment.
In this regard, we note the anomalous Hall effect \cite{2010_Nagaosa_AnomalousHallEffect} is responsible for most Hall sensor voltage measurements involving chiral molecules \cite{2024_Bloom_ChiralInducedSpinSelectivity} and has been observed for monolayers of chiral molecules \cite{2017_BenDor_MagnetizationSwitchingFerromagnets}.
Additionally, the anomalous Hall effect has been observed using Hall sensors fabricated with GaN \cite{2011_Yin_ObservationPhotoinducedAnomalous}, which was the sensor material chosen by the reference experiment based on its long spin lifetime \cite{2017_Kumar_ChiralityInducedSpinPolarizationPlacesSymmetry}.

Modelling the anomalous Hall effect is beyond the scope of this work. Accordingly, we have reported our results as magnetic fields, which are relevant to the ordinary Hall effect, rather than as voltages, thereby eliminating the dependence on the unknown $R$ and $C$ values from the experiment.
This approach does not preclude qualitative comparisons between trends in $\widetilde{V}_{\mathrm{H,peak}}$ and $\bar{B}^z_{\mathrm{H,peak}}$ with changing monolayer properties, as both $\widetilde{V}_{\mathrm{H,peak}}$ and $\bar{B}^z_{\mathrm{H,peak}}$ systematically quantify the size of the spin polarization throughout the monolayer, even if the anomalous Hall effect contributes to the experimental Hall voltage.

Detailed investigation into all listed points of discrepancy, including the anomalous Hall effect, would be required to seek quantitative agreement with experiment.
As a first step, we only aim to show that qualitative, signature trends can be captured from our model, although in \refsec{}~\ref{Sec::ParameterSweep}, we do consider parameter variations that approximate the effect of interactions.

\subsection{Possible impact of current on magnetic-field results}\label{Sec::Current}

\begin{figure}[htbp!]
    \includegraphics{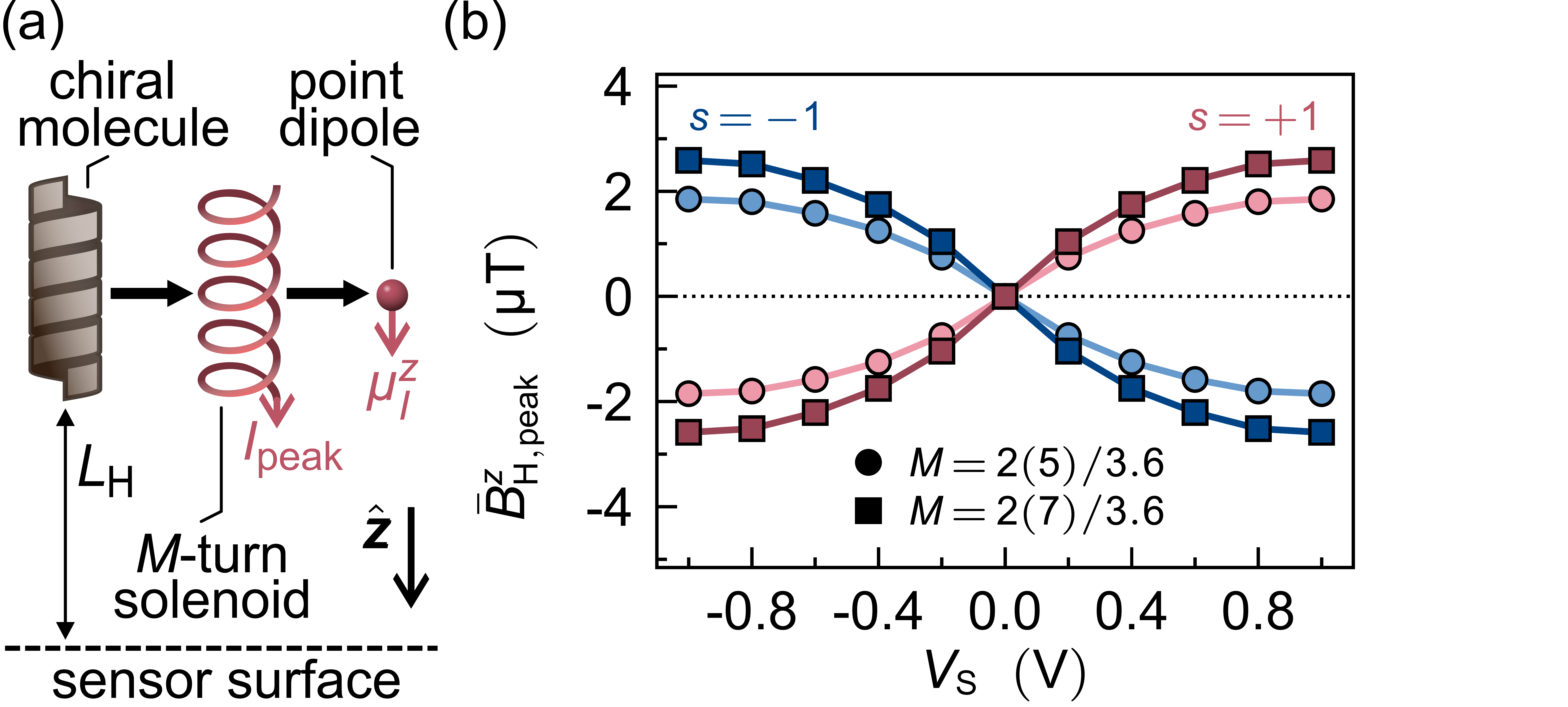}
    \caption{
        (a) Diagram (not to scale) showing treatment of each chiral molecule in our monolayer model as an $M$-turn solenoid with a current $I_{\text{peak}}$.
        Each solenoid is subsequently approximated as a point dipole with magnetic moment $\mu^z_{I} = s M\pi r^2 I_{\text{peak}}$.
        (b) Peak magnetic field from the peak current in our monolayer model, which was calculated by treating each chiral molecule as a point dipole according to (a), for various $M$, $s$, and $V_{\text{S}}$.
        Additional parameters are given in \refsec{}~\ref{Sec::Parameters}.
    }
    \label{Fig::Current}
\end{figure}

To the best of our knowledge, the possible impact of current in a monolayer of chiral molecules on Hall sensor measurements that intend to detect a spin polarization has not yet been considered.
To contextualize our $\bar{B}^z_{\mathrm{H,peak}}$ results from \reffig{}~\ref{Fig::Experiment}(f), which arise from the spin polarization, we estimate the peak magnetic field arising from current in our monolayer model.
We assume that the peak magnetic field from current follows from the temporal peak of the aggregate change in charge per unit time throughout the molecule used in our monolayer model, denoted $I_{\text{peak}}$.
Thus, we take $I_{\text{peak}}$ as the value of $-e\,\Delta n(t) / \Delta t$ at the time of maximal magnitude, where $\Delta n(t) = \sum_{m=1}^{\mathbb{M}} [n_m(t + \Delta t) - n_m(t)]$ and $\Delta t$ is the simulation timestep.
A plot of $-e\,\Delta n(t) / \Delta t$ is shown in \reffig{}~S12 in \refsec{}~S9 of \refsuppinfo{}, which closely approximates the current entering the molecule.
To simplify the magnetic-field calculation, we approximate each chiral molecule, effectively a solenoid, as a point dipole located at the center of the corresponding molecule, as shown in \reffig{}~\ref{Fig::Current}(a).
This approximation is reasonable because a point dipole with a magnetic moment $\bm{\mu}_I = \mu^z_I \hat{\bm{z}}$, where $\mu^z_I = s M \pi r^2 I_{\mathrm{peak}}$, generates a magnetic flux through the Hall sensor surface that is within 10\% of that generated by a solenoid carrying a current $I_{\text{peak}}$, as shown in \refsec{}~S9 of \refsuppinfo{}.
To calculate $\bar{B}^z_{\mathrm{H,peak}}$ from $I_{\text{peak}}$ in every molecule of the monolayer model, we used \refeq{}~\refeqnum{Eq::DipoleSurface_Average} with $\mu^z_{m}(t) \rightarrow \mu^z_I$ and $L_m \rightarrow L_{\text{H}} + l(\mathbb{M}-1)/2N$.
\Reffig{}~\ref{Fig::Current}(b) shows results for various monolayers at different $V_{\mathrm{S}}$. 

The similar trends in \reffigs{}~\ref{Fig::Experiment}(f) and \ref{Fig::Current}(b) combined with the much larger magnetic-field magnitude of the latter emphasizes the importance of considering the possible impact of current during magnetic-field measurements.
In the next section, we show that parameter variations can increase $|\bar{B}^z_{\mathrm{H,peak}}|$ from the spin polarization and decrease $|\bar{B}^z_{\mathrm{H,peak}}|$ from current, such that they are similar magnitudes.

\subsection{Impact of transport properties on magnetic-field results}\label{Sec::ParameterSweep}

\begin{figure*}[htbp!]
    \includegraphics{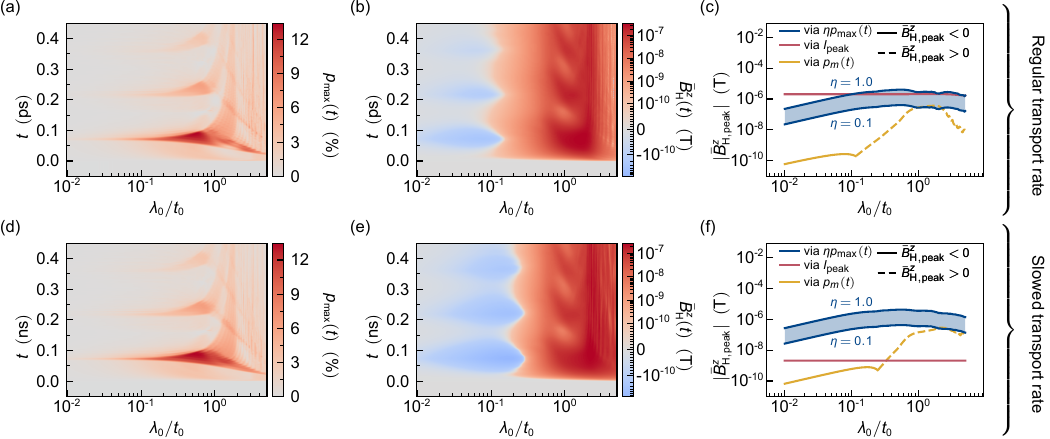}
    \caption{
        (a) $\lambda_0/t_0$ dependence of the maximum spin polarization at any site within a one-lead chiral molecule.
        (b) $\lambda_0/t_0$ dependence of the magnetic field from the spin polarization calculated for our monolayer model, depicted in \reffig{}~\ref{Fig::Experiment}(d), with the peak magnetic field indicated.
        (c) $\lambda_0/t_0$ dependence of the peak magnetic field calculated for our monolayer model from the spin polarization (yellow), from the peak current (red), and from a uniform spin polarization $p_m(t) = \eta p_{\text{max}}(t)$, with $0.1\leq\eta\leq1.0$ (blue).
        The transient calculations used for (a), (b), and (c) apply $V_{\text{S}}$ at $t=0$ and use a fixed $t_0=0.2\;\text{eV}$ and $\Gamma_0=0.05\;\text{eV}$.
        Additional parameters are given by \refsec{}~\ref{Sec::Parameters}.
        Adjusted results for the $\lambda_0/t_0$ dependence of the (d) maximum spin polarization, (e) magnetic field, and (f) peak magnetic field, now with $t_0=0.2\;\text{meV}$ and $\Gamma_0=0.05\;\text{meV}$ such that the transport timescale is slowed by a factor of 1000.
    }
    \label{Fig::Interactions}
\end{figure*}

Although the TDLB formula has limited our analysis to noninteracting systems, we can estimate the effect of interactions within molecules by increasing the SOC parameter \cite{2020_Ghazaryan_AnalyticModelChiralInducedSpin}.
Then, \reffig{}~\ref{Fig::Interactions}(a) shows the spatial-maximum spin polarization within a one-lead molecule at a given time, denoted $p_{\mathrm{max}}(t)$, as a function of $\lambda_0/t_0$, with $t_0$ fixed.
\Reffig{}~\ref{Fig::Interactions}(b) shows $\bar{B}^z_{\mathrm{H}}(t)$ from the spin polarization in the monolayer as a function of $\lambda_0/t_0$, from which $\bar{B}^z_{\mathrm{H,peak}}$ was extracted and plotted in \reffig{}~\ref{Fig::Interactions}(c).
When $\lambda_0/t_0$ is sufficiently small, the maximum spin polarization increases with increasing $\lambda_0/t_0$.
This trend continues until $p_{\mathrm{max}}(t)$ reaches a maximum of 13.4\% when $\lambda_0/t_0 \approx 0.5$.
As $\lambda_0/t_0$ increases, the reflection process becomes more complicated than that described in \refsec{}~\ref{Sec::OneLead}.
Eventually, the maximum spin polarization decreases with increasing $\lambda_0/t_0$.
Similar trends follow for the magnitude of $\bar{B}^z_{\mathrm{H,peak}}$ from the spin polarization.
\Reffig{}~\ref{Fig::Interactions}(c) also shows $\bar{B}^z_{\mathrm{H,peak}}$ estimated from current in the monolayer, which is largely independent of $\lambda_0/t_0$ because the average group velocity of both spin states is unaffected by $\lambda_0$, as shown in \reffig{}~\ref{Fig::OneLead}(e).

The logarithmic dependence of \refeq{}~\refeqnum{Eq::DipoleSurface_Average} on $L_m$ indicates that $\bar{B}^z_{\mathrm{H}}(t)$ is 
only weakly dependent on the distance between each sublayer and the sensor surface.
Thus, any sublayers that have opposing spin polarizations partially cancel when contributing to the total magnetic field.
Based on the opposing spin polarizations present throughout the molecule shown in \reffig{}~\ref{Fig::OneLead}(d), this cancellation occurs in our magnetic-field results.
In contrast, if each sublayer had a spin polarization with the same sign, the magnetic field could be dramatically larger.
For example, suppose that the spin polarization is uniform such that $p_m(t) = \eta p_{\mathrm{max}}(t)$ for each site $m$, where $0 < \eta \leq 1$, the resulting $|\bar{B}^z_{\mathrm{H,peak}}|$ shown in \reffig{}~\ref{Fig::Interactions}(c) is much larger than that of the unmodified spin-polarization distribution.
A transport calculation that uses the same tight-binding model considered here, but includes interactions within the molecule, resulted in a spin polarization with a consistent sign and similar magnitude at each atomic site \cite{2024_Chiesa_ManyBodyModelsChiralityInducedSpin}.
This characteristic suggests that interactions might lead to a more uniform spin-polarization distribution, thus increasing the resulting magnetic field.

To estimate the effect of interactions between molecules, which is expected to increase transport timescales, we decrease $t_0$ and $\Gamma_0$.
For a given $\lambda_0/t_0$ ratio, decreasing $t_0$ and $\Gamma_0$ by a factor simply increases the transport timescale by that same factor.
This increased timescale is illustrated by \reffigs{}~\ref{Fig::Interactions}(d) and \ref{Fig::Interactions}(e), which plot $p_{\mathrm{max}}(t)$ and $\bar{B}^z_{\mathrm{H}}(t)$, respectively, using values for $t_0$ and $\Gamma_0$ that are decreased compared to those used in \reffigs{}~\ref{Fig::Interactions}(a) and \ref{Fig::Interactions}(b).
An increased timescale reduces the $|\bar{B}^z_{\mathrm{H,peak}}|$ estimated from current, but leaves the $|\bar{B}^z_{\mathrm{H,peak}}|$ from the spin polarization unchanged.
\Reffig{}~\ref{Fig::Interactions}(f) shows that a $1000\times$ decrease in the $t_0$ and $\Gamma_0$ values used in \reffig{}~\ref{Fig::Interactions}(c) causes the $\bar{B}^z_{\mathrm{H,peak}}$ from the spin polarization to have a magnitude similar to that estimated from current when $\lambda_0/t_0 \approx 0.3$.
Ultimately, \reffigs{}~\ref{Fig::Interactions}(c) and \ref{Fig::Interactions}(f) suggest that the presence of interactions could significantly increase the contribution of the spin polarization to the magnetic field, relative to that of current.

\section{Conclusions}\label{Sec::Conclusions}

By performing time-dependent calculations for a chiral molecule represented by a tight-binding model, we conclude the following:
(1) The presence of chirality in a tight-binding model can cause an asymmetry in the group velocity of electrons with different spin states.
(2) This group-velocity asymmetry leads to a nonzero spin polarization (based on occupancy, as we have defined it) in the transient, even in the absence of interactions.
(3) When one lead is present and no interactions are present, the spin polarization must decay to zero in the steady state because no net electron transport occurs.
(4) When two leads are present at different biases, net transport facilitates a nonzero spin polarization in the steady state, even in the absence of interactions.
(5) Increasing the SOC strength relative to other couplings in the one-lead (two-lead) molecule can increase the maximum-transient (steady-state) spin polarization, but decreasing other couplings increases the time required to reach the maximum-transient (steady-state) value.
(6) Our one-lead spin-polarization results qualitatively reproduce trends in the experimental magnetic fields observed for monolayers with varying molecule length, chirality, and driving voltage \cite{2017_Kumar_ChiralityInducedSpinPolarizationPlacesSymmetry}.
(7) Based on trends in the magnetic field estimated from the peak current in our model, it is important for experiments to consider the possible impact of current on magnetic fields attributed to a spin polarization.
Ultimately, these outcomes show that an asymmetry in the group velocity of electrons with different spin states is a possible contributor to experimental magnetic-field signatures arising from the CISS effect in dynamic (transient) conditions.
\vspace{0.5cm plus 1ex minus 0.2ex}

\nocite{2013_Tuovinen_TimeDependentLandauer-ButtikerFormulaTransient,2014_Tuovinen_TimeDependentLandauerButtikerFormulaApplicationa,2016_Tuovinen_TimeDependentLandauer-buttikerFormalismSuperconducting,1972_Kolbig_ProgramsComputingLogarithm,2008_Michel_FastComputationGauss,2019_Tuovinen_DistinguishingMajoranaZeroModes,2024__MathematicaVer142WolframResearch,2023__MpmathPythonLibrary,2017_Kumar_ChiralityInducedSpinPolarizationPlacesSymmetry,2009_Feller_IntroductionProbabilityTheory,2009_Votyakov_AnalyticModelsHeterogenousMagnetic,1994_Mantorov_InterpretationHallEffectMeasurements,1998_Heras_ExplicitExpressionsElectric,2020_Ortner_MagpylibFreePythonPackage}

\begin{acknowledgments}
This work was supported in part by the Natural Sciences and Engineering Research Council of Canada (NSERC) and in part by Alberta Innovates.
J.F. acknowledges support from Olle Engkvists Stiftelse.
\end{acknowledgments}

\bibliography{ref.bib}

\end{document}

\title{Supplemental material:\\Spin dependence of charge dynamics and group velocity in chiral molecules}

\author{Riley Stuermer}
\author{Collin VanEssen}
\author{Jacob Byers}
\author{Keith Ferrer}
\affiliation{Department of Electrical and Computer Engineering, University of Alberta, Edmonton, Alberta T6G 2V4, Canada}
\author{Prasad Gudem}
\affiliation{Department of Electrical and Computer Engineering, University of California at San Diego, La Jolla, California 92093, USA}
\author{Diego Kienle}
\affiliation{81245 Munich, Germany}
\author{Jonas Fransson}
\affiliation
{Department of Physics and Astronomy, Uppsala University, 75120 Uppsala, Sweden}
\author{Mani Vaidyanathan}
\email[Contact author: ]{maniv@ualberta.ca}
\affiliation{Department of Electrical and Computer Engineering, University of Alberta, Edmonton, Alberta T6G 2V4, Canada}

\maketitle

\tableofcontents

\section{Optimization of time-dependent Landauer--B{\"u}ttiker formula at finite temperature}

The time-dependent Landauer--B{\"u}ttiker (TDLB) formula \cite{2013_Tuovinen_TimeDependentLandauer-ButtikerFormulaTransient,2014_Tuovinen_TimeDependentLandauerButtikerFormulaApplicationa} has been expressed in terms of analytic functions at nonzero temperature \cite{2016_Tuovinen_TimeDependentLandauer-buttikerFormalismSuperconducting}.
Of the three relevant functions, $\Lambda_{\alpha,jk}$ and $\Omega_{\alpha,jk}$ can be written in terms of digamma functions, whereas $\Pi_{\alpha,jk}(t)$ has been expressed in terms of hypergeometric functions.
Efficient algorithms exist for computing digamma functions \cite{1972_Kolbig_ProgramsComputingLogarithm} and hypergeometric functions \cite{2008_Michel_FastComputationGauss}.
However, because $\Pi_{\alpha,jk}(t)$ is time-dependent, it must be calculated at each simulation timestep.
In contrast, $\Lambda_{\alpha,jk}$ and $\Omega_{\alpha,jk}$ can be calculated once and used for all simulation timesteps.
These characteristics make calculating $\Pi_{\alpha,jk}(t)$ the rate-limiting step of the TDLB formula, which motivates optimizing its implementation.

% #NOTE: In mathematica, the computation speed of the hypergeometric function is slow compared to that of the digamma function.
% We have not tested if this holds in general.

As an intermediate step to defining $\Pi_{\alpha,jk}(t)$ in terms of hypergeometric functions, $\Pi_{\alpha,jk}(t)$ can be formulated as an infinite series according to \cite{2019_Tuovinen_DistinguishingMajoranaZeroModes}
\begin{equation}\label{Eq::Pi_Sum}
    \begin{split}
        \Pi_{\alpha,jk}(t) ={}& i \frac{e^{-i(\epsilon_j - \epsilon^*_k )t/\hbar} f(\epsilon^*_k - \mu_{\alpha})}{(\epsilon^*_k-\epsilon_j)(\epsilon^*_k-\epsilon_j-E_{\alpha})} + \frac{e^{-2 \pi k_{\text{B}} T \Phi_{\alpha}(\epsilon_j) t/\hbar}}{(2 \pi)^3(k_{\text{B}} T)^2} \\
        &\times \sum^{\infty}_{\nu=0} \frac{e^{-2 \pi k_{\text{B}} T \nu t/\hbar}}{[\nu+\Phi_{\alpha}(\epsilon_j)][\nu+\Phi_{\alpha}(\epsilon^*_k)][\nu+\Phi(\epsilon_j)]}\eqcomma{}
    \end{split}
\end{equation}
where $\mu_{\alpha} = \mu + E_{\alpha}$, $\Phi(E) = \frac{1}{2} - \frac{E - \mu}{2 \pi i k_{\text{B}} T}$, $\Phi_{\alpha}(E) = \Phi(E-E_{\alpha})$, and $f(E) = ( \exp \frac{E}{k_{\text{B}} T} + 1)^{-1}$ is the Fermi function.
Other symbols have the same meaning as specified in the main text.
The magnitude of terms in the sum of \refeq{}~\refeqnum{Eq::Pi_Sum} decrease with increasing $\nu$, and the series converges as $\nu \rightarrow \infty$.
The rate of convergence of the series increases with increasing electron temperature $T$ or simulation time $t$.

\begin{figure}[htbp!]
    \captionsetup[subfloat]{farskip=0pt,captionskip=0pt}
    \subfloat[][]{\includegraphics{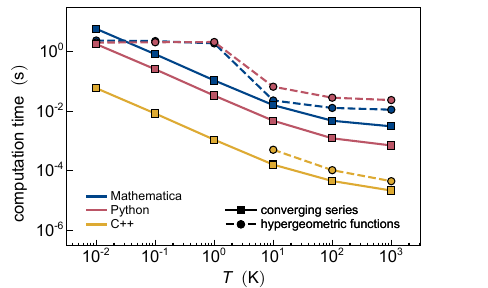}\label{Fig::TemperatureDependentComputationCost_TemperatureSweep}}\\
    \subfloat[][]{\includegraphics{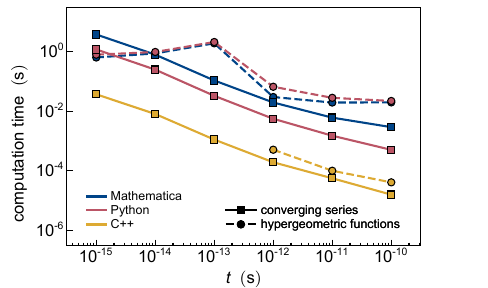}\label{Fig::TemperatureDependentComputationCost_TimeSweep}}
    \caption{
        Time to compute $\bm{\Pi}_{\alpha}(t)$ for (a) various $T$ at a fixed $t=0.1\;\text{ps}$ and (b) various $t$ at a fixed $T=1\;\text{K}$.
    }
\end{figure}

Using both \refeq{}~\refeqnum{Eq::Pi_Sum} and the hypergeometric counterpart, we measured the time required to compute the matrix $\bm{\Pi}_{\alpha}(t)$ for a system formulated according to the tight-binding model for a one-lead chiral molecule outlined in the main text.
The model parameters used were identical to those from the main text, except here $N=3$ and $M=2$.
The resulting computation times for various $T$ and $t$ are shown in \reffigs{}~\ref{Fig::TemperatureDependentComputationCost_TemperatureSweep} and \ref{Fig::TemperatureDependentComputationCost_TimeSweep}, respectively. 
Implementations in Mathematica \cite{2024__MathematicaVer142WolframResearch}, Python, and C++ were considered.
When calculating $\Pi_{\alpha,jk}(t)$ according to \refeq{}~\refeqnum{Eq::Pi_Sum}, we included terms until the successive term contributed less than $10^{-6}\%$ to $\Pi_{\alpha,jk}(t)$.
Tightening this tolerance would improve the precision of the calculation at the expense of an increased computation time.
When calculating $\Pi_{\alpha,jk}(t)$ using hypergeometric functions, the built-in hypergeometric function was used in Mathematica, the hypergeometric function from the mpmath library \cite{2023__MpmathPythonLibrary} was used in Python, and a program that efficiently calculates the hypergeometric function \cite{2008_Michel_FastComputationGauss} was used in C++.
This C++ program only yields accurate results for the hypergeometric function within a restricted range of arguments \cite{2008_Michel_FastComputationGauss}, thereby limiting its applicability to a subset of the $T$ and $t$ that we considered.

Ultimately, for the majority of $T$ and $t$ considered, calculating $\Pi_{\alpha,jk}(t)$ according to the converging series of \refeq{}~\refeqnum{Eq::Pi_Sum} could improve the computation time compared to calculations relying on hypergeometric functions.
However, for low temperatures or small simulation times, i.e., when the series converges slowly, implementations using hypergeometric functions may still be beneficial.

\section{Analytical group velocity of chiral molecules}

Starting from the tight-binding Hamiltonian for a chiral molecule given in the main text, Schr{\"o}dinger's equation for atomic site $m$ can be written as
\begin{equation}
    \begin{split}
        E \phi_m =&{} \begin{bmatrix} \varepsilon_0 & 0 \\ 0 & \varepsilon_0 \end{bmatrix} \phi_{m} - \begin{bmatrix} t_0 & 0 \\ 0 & t_0 \end{bmatrix} (\phi_{m+1} + \phi_{m-1}) \\
        &+ \begin{bmatrix} i \lambda_0 \chi^z & 0 \\ 0 & -i \lambda_0 \chi^z \end{bmatrix} (\phi_{m+2} -  \phi_{m-2})\eqcomma{}
    \end{split}
\end{equation}
where $\phi_{m} = [\phi_{m,+}, \phi_{m,-}]^T$ is an expansion coefficient spinor for the basis function at atomic site $m$, $E$ is the energy eigenvalue, and other symbols have the same meaning as specified in the main text.
Like in the main text, we have neglected the $\hat{\bm{x}}$ and $\hat{\bm{y}}$ components of $\bm{\chi}_m$, which allows for separable equations for spin $|+\rangle$ and $|-\rangle$.
Supposing $\phi_m = A e^{i k d m}$, where $k$ is the wave vector, an $E$-$k$ relation follows as
\begin{equation}\label{Eq::AnalyticalDispersianRelation}
    E_{\pm}(k) = \varepsilon_0 - 2 t_0 \cos{k d} \mp 2 \lambda_0 \chi^z \sin{2 k d}\eqperiod{}
\end{equation}
The group velocity is given by 
\begin{equation}\label{Eq::AnalyticalGroupVelocity}
    v_{g,\pm}(k) = \frac{1}{\hbar}\,\frac{\text{d}}{\text{d} k} E_{\pm}(k)
    = \frac{2 t_0 d}{\hbar} \sin{k d} \mp \frac{4 \lambda_0 \chi^z d}{\hbar} \cos{2 k d}\eqcomma{}
\end{equation}
which has a maximum magnitude when $k$ satisfies
\begin{equation}\label{Eq::AnalyticalGroupVelocity_MaximumCondition}
    \frac{\text{d}}{\text{d} k} v_{g,\pm}(k) = \frac{2 t_0 d^2}{\hbar} \cos{k d} \pm \frac{16 \lambda_0 \chi^z d^2}{\hbar} \cos{k d} \, \sin {k d} = 0\eqperiod{}
    % #NOTE: t_0 \cos{k d} = \mp 8 \lambda_0 \chi^z \cos{k d} \sin {k d} = 0
\end{equation}
Assuming that the chiral molecule parameters satisfy $| t_0 | > | 8 \lambda_0 \chi^z |$ (which corresponds to weak spin-orbit coupling) the only solutions to \refeq{}~\refeqnum{Eq::AnalyticalGroupVelocity_MaximumCondition} occur at $k=\pm\frac{\pi}{2d}$.
The energies and group velocities at these $k$ values are given by
\begin{equation}
    E_{\pm}\Big(+\frac{\pi}{2d}\Big)= E_{\pm}\Big(-\frac{\pi}{2d}\Big) = \varepsilon_0
\end{equation}
and
\begin{equation}\label{Eq::SelectedGroupVelocity}
    \begin{split}
        v_{g,\pm}\Big(+\frac{\pi}{2d}\Big) &= \frac{2 d}{\hbar} (+t_0 \pm 2 \lambda_0 \chi^z) \\
        v_{g,\pm}\Big(-\frac{\pi}{2d}\Big) &= \frac{2 d}{\hbar} (-t_0 \pm 2 \lambda_0 \chi^z)\eqcomma{}
    \end{split}
\end{equation}
respectively.
Because we have assumed $| t_0 | > | 8 \lambda_0 \chi^z | > | 2 \lambda_0 \chi^z |$, the sign of $v_{g,\pm}$ in \refeq{}~\refeqnum{Eq::SelectedGroupVelocity} is determined entirely by the term containing $t_0$.
Separating \refeq{}~\refeqnum{Eq::SelectedGroupVelocity} into $v_{g,\pm} > 0$ and $v_{g,\pm} < 0$ leads to
\begin{equation}\label{Eq::SelectedGroupVelocity_Simplfied}
    v_{g,\pm} = \frac{2d}{\hbar}\begin{cases}
        +|t_0| \pm 2 \lambda_0 s |\chi^z|, & v_{g,\pm} > 0 \\
        -|t_0| \pm 2 \lambda_0 s |\chi^z|, & v_{g,\pm} < 0
    \end{cases}\eqcomma{}
\end{equation}
which is the result from the main text.

\begin{figure*}[htbp!]
    \captionsetup[subfloat]{farskip=0pt,captionskip=0pt}
    \subfloat[][]{\includegraphics{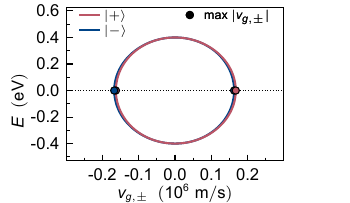}\label{Fig::GroupVelocityVisualization_LowSOC}}
    \subfloat[][]{\includegraphics{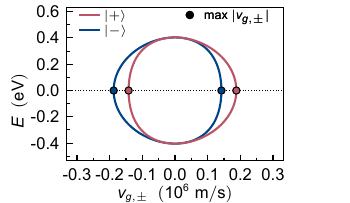}\label{Fig::GroupVelocityVisualization_MediumSOC}}
    \subfloat[][]{\includegraphics{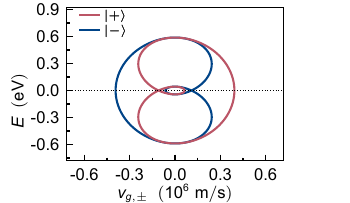}\label{Fig::GroupVelocityVisualization_HighSOC}}
    \\
    \subfloat[][]{\includegraphics{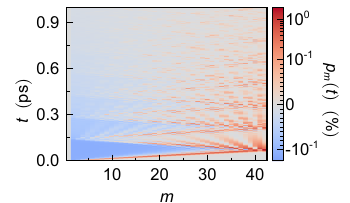}\label{Fig::GroupVelocityVisualization_LowSOC_SpinPolarization}}
    \subfloat[][]{\includegraphics{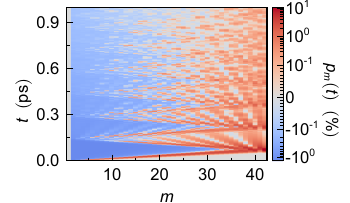}\label{Fig::GroupVelocityVisualization_MediumSOC_SpinPolarization}}
    \subfloat[][]{\includegraphics{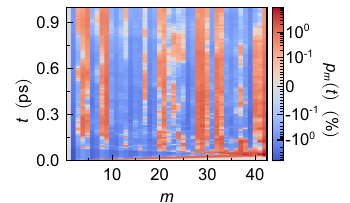}\label{Fig::GroupVelocityVisualization_HighSOC_SpinPolarization}}
    \caption{
        Analytical $E$-$v_g$ plots with (a) $\lambda_0=5\;\text{meV}$, (b) $\lambda_0=50\;\text{meV}$, and (c) $\lambda_0=500\;\text{meV}$.
        Other parameters are given by the main text.
        (d), (e), and (f) Numerical spin polarization for a one-lead chiral molecule obtained with parameters corresponding to (a), (b), and (c), respectively.
    }
    \label{Fig::GroupVelocityVisualization}
\end{figure*}

\Reffig{}~\ref{Fig::GroupVelocityVisualization} shows analytic plots of $v_{g,\pm}$ as a function of energy for different spin-orbit coupling values and the corresponding numerical spin polarization.
The $E$-$v_{g}$ plots were obtained by reparameterizing \refeq{}~\refeqnum{Eq::AnalyticalGroupVelocity} in terms of $E$ with \refeq{}~\refeqnum{Eq::AnalyticalDispersianRelation}.
As the spin-orbit coupling is increased beyond the point where \refeq{}~\refeqnum{Eq::SelectedGroupVelocity_Simplfied} is valid, the reflection pattern becomes complicated, as shown by \reffig{}~\ref{Fig::GroupVelocityVisualization_HighSOC_SpinPolarization}.

% #NOTE: An even larger SOC results in a decreased spin polarization as the two circles overlap.
% This is consistent with (and explains) the finding from Chiesa where spin polarization disappears when only one coupling term is included (i.e., one is negligible compared to another).

\section{Schematic of electron-beam reflection process}

To illustrate the reflection process resulting from the superposition of electron beams described in the main text, we construct a phenomenological schematic.
The density and velocity of the spin-$|\pm\rangle$ beam after the $i$-th reflection are denoted $n_{\pm,i}$ and $v_{g,\pm,i}$, respectively.
The total density of the superposition of beams in the steady state is $n_{\pm} = \sum_{i=0}^{\infty} n_{\pm,i}$, which has a corresponding spin polarization $p = (n_{+} - n_{-})/(n_{+} + n_{-})$.
We calculate these quantities according to the following rules:
\begin{enumerate}
    \item From \refeq{}~\refeqnum{Eq::SelectedGroupVelocity_Simplfied}, electron beams with different spin have different group velocities based on the direction of travel according to
    \begin{equation}
        v_{g,\pm,i} = \begin{cases}
        +v_g \pm \Delta v_g > 0, & \text{even } i \\
        -v_g \pm \Delta v_g < 0, & \text{odd } i
        \end{cases}\eqcomma{}
    \end{equation}
    where $v_g > \Delta v_g > 0$ are parameters for the group velocity, and the sign of $v_{g,\pm,i}$ determines the direction of travel.
    \item A source lead continually injects spin beams with fixed densities given by
    \begin{equation}\label{Eq::DensityFlipping}
        n_{\pm,0} = n \frac{v_g}{|v_{g,\pm,0}|}\eqcomma{}
    \end{equation}
    where $n$ is a parameter for the injected density. The scaling on the right-hand side of \refeq{}~\refeqnum{Eq::DensityFlipping} ensures the injected electron flux is equal for both spins $|v_{g,\pm,0}| n_{\pm,0} = v_g n$.
    \item Any change in group velocity following reflection from a boundary is accompanied by a counteracting change in density according to
    \begin{equation}\label{Eq::DensityScaling}
        n_{\pm,i+1} = n_{\pm,i} \frac{|v_{g,\pm,i}|}{|v_{g,\pm,i+1}|}
    \end{equation}
    such that the magnitude of the electron flux is conserved $|v_{g,\pm,i+1}| n_{\pm,i+1} = |v_{g,\pm,i}| n_{\pm,i}$ (assuming there are no other losses).
    \item A source lead is assumed to be connected, but a drain lead can optionally be connected at the other end of the system.
    If a beam reflects at a boundary of the system that has a lead, the density of the reflected beam is decreased to account for losses owing to transmission, which amounts to multiplying the right-hand side of \refeq{}~\refeqnum{Eq::DensityScaling} by $1-T$, where $0<T<1$.
    If a beam reflects at a boundary of the system that has no lead, no losses occur, and \refeq{}~\refeqnum{Eq::DensityScaling} is unadjusted.
\end{enumerate}

\begin{figure}[htbp!]
    \captionsetup[subfloat]{farskip=0pt,captionskip=0pt}
    \subfloat[][]{\includegraphics{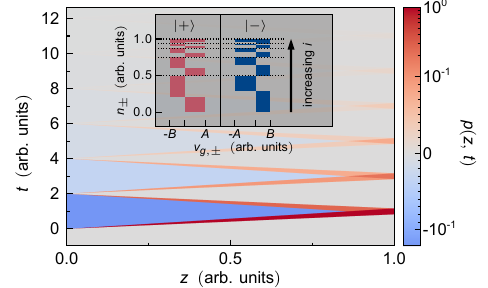}\label{Fig::SpinPlarizationVisualization_OneLead}}\\
    \subfloat[][]{\includegraphics{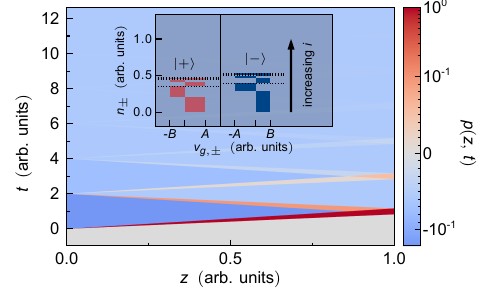}\label{Fig::SpinPlarizationVisualization_TwoLeads}}
    \caption{
        Phenomenological schematic of spin polarization for a chiral molecule with (a) one lead and 
        (b) two leads.
        Insets show the density and group velocity of each spin beam involved in the reflection process.
        We denote $|v_g+\Delta v_g| = A$ and $|v_g-\Delta v_g| = B$.
    }
    \label{Fig::SpinPlarizationVisualization}
\end{figure}

Resulting schematics for the one-lead and two-lead case are shown in \reffigs{}~\ref{Fig::SpinPlarizationVisualization_OneLead} and \ref{Fig::SpinPlarizationVisualization_TwoLeads}, respectively, which qualitatively match the numerically calculated spin polarization presented in the main text. 
This agreement corroborates our description of the reflection process.

% #NOTE: The superposition of all electron beams following infinite reflections converges to a steady-state electron density.
% In the one-lead case, we normalize the steady-state electron density to 2, analogous to complete filling of all atomic sites.
% We use the initial densities corresponding to the normalization from the one-lead case for the two-lead case.
% Then, the steady-state electron density converges to $\approx 1$ in the two-lead case, which is consistent with equal transmission for the source and drain leads.

The group velocity and density of each spin-$|+\rangle$ (spin-$|-\rangle$) beam is given by the red (blue) rectangles shown in the insets of \reffig{}~\ref{Fig::SpinPlarizationVisualization}.
The area of each rectangle gives the flux of the corresponding spin beam, which are tabulated in Table~\ref{Tab::ElectronBeamFlux}.
The flux of beam $i$ is spin-independent.
In the two-lead case, losses at the drain breaks the flux symmetry between incident and reflected beams, which ultimately results in the density differences that culminate in a nonzero spin polarization in the steady state.

\begin{table}
    \caption{Electron flux (arb. units) of beams from \reffig{}~\ref{Fig::SpinPlarizationVisualization}}
    \label{Tab::ElectronBeamFlux}
    \begin{ruledtabular}
    \begin{tabular}{ccccc}
    &\multicolumn{2}{c}{one lead} & \multicolumn{2}{c}{two leads} \\
    $i$ & $|+\rangle$ & $|-\rangle$ & $|+\rangle$ & $|-\rangle$ \\[3pt] \hline
    $0$ & $+1$ & $+1$ & $+1$ & $+1$ \\
    $1$ & $-1$ & $-1$ & $-1/2$ & $-1/2$ \\
    $2$ & $+1/2$ & $+1/2$ & $+1/4$ & $+1/4$ \\
    $3$ & $-1/2$ & $-1/2$ & $-1/8$ & $-1/8$ \\
    $4$ & $+1/4$ & $+1/4$ & $+1/16$ & $+1/16$ \\
    $5$ & $-1/4$ & $-1/4$ & $-1/32$ & $-1/32$ \\
    \vdots & \vdots & \vdots & \vdots & \vdots \\
    \end{tabular}
\end{ruledtabular}
\end{table}

\section{Symmetry of spin-resolved currents}\label{Sec::SymmetryCurrent}

\begin{figure}[htbp!]
    \captionsetup[subfloat]{farskip=0pt,captionskip=0pt}
    \subfloat[][]{\includegraphics{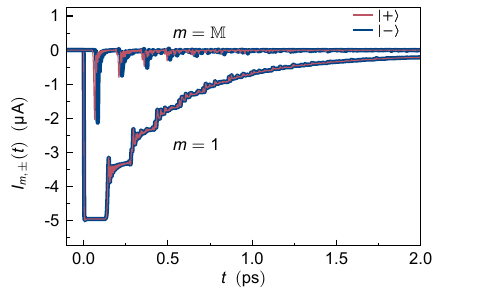}\label{Fig::CurrentConservation_OneLead}}\\
    \subfloat[][]{\includegraphics{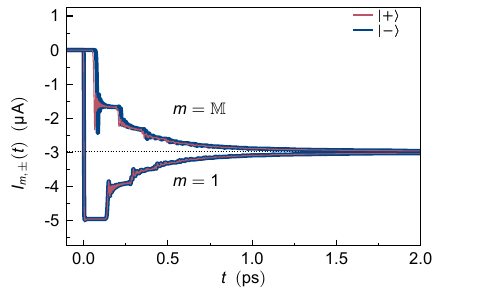}\label{Fig::CurrentConservation_TwoLeads}}
    \caption{
        Spin-resolved currents $I_{m,\pm}(t)$ associated with site 1 and $\mathbb{M}$ for (a) the one-lead case and (b) the two-lead case.
        All parameters were set according to the main text, except for $\lambda_0$, which was increased to $50\;\text{meV}$ so that a slight delay between spin currents is visible for site $\mathbb{M}$.
    }
    \label{Fig::CurrentConservation}
\end{figure}

The current through a cross section of our molecule can be calculated using the off-diagonal components of the time-dependent one-particle reduced density matrix described by the main text \cite{2014_Tuovinen_TimeDependentLandauerButtikerFormulaApplicationa}.
We consider the current between site 1 and 2 (which includes hopping between site 1 and 3) and the current between site $\mathbb{M}-1$ and $\mathbb{M}$ (which includes hopping between site $\mathbb{M}-2$ and $\mathbb{M}$).
The resulting currents, which are each proportional to an electron flux, are shown in \reffigs{}~\ref{Fig::CurrentConservation_OneLead} and \ref{Fig::CurrentConservation_TwoLeads} for the one-lead and two-lead case, respectively.

Based on \reffig{}~\ref{Fig::CurrentConservation}, at site 1, the spin fluxes coincide at any time $t$. At site $\mathbb{M}$, the spin-$|-\rangle$ flux is slightly delayed compared to the spin-$|+\rangle$ flux, which follows from the group-velocity asymmetry of electrons with different spin states.
In the steady state, the fluxes at either end of the molecule converge to the same spin-independent value.
In the one-lead case, the steady-state flux is zero.
In the two-lead case, the steady-state flux is nonzero, which is consistent with the presence of net transport.

As defined by the main text, a residual $+\hat{\bm{z}}$-moving flux is a net flux, i.e., any portion of a $+\hat{\bm{z}}$-moving flux that is not countered by a $-\hat{\bm{z}}$-moving flux.
This net flux is the quantity shown in \reffig{}~\ref{Fig::CurrentConservation_TwoLeads}, which thus shows that the residual $+\hat{\bm{z}}$-moving fluxes of the spin-$|+\rangle$ and spin-$|-\rangle$ beams are equal (up to a small delay).

\section{Use of expectation values to calculate magnetic field from monolayer}

Our magnetic-field calculation from the main text relies on the expectation value of electrons, which is not necessarily the situation relevant to an experimental magnetic-field measurement where each electron could occupy a particular state throughout time.
Here, to justify our use of expectation values, we show that the results obtained by the reference experiment approximate an expectation value.

For an experimental trial $i$ that measures the Hall voltage resulting from a monolayer of chiral molecules, the dipole at atomic site $m$ in molecule $n$ has a magnetic dipole moment $\mu^z_{m,n,i}(t)$.
In general, this dipole has a contribution to the Hall voltage $V_{\text{H},i}(t)$ that depends on a weighting coefficient $w_{m,n}$ such that
\begin{equation}\label{Eq::HallSum_Trial}
    V_{\text{H},i}(t) = \sum^{\mathbb{M}}_{m=1} \sum^{\mathbb{N}}_{n=1} w_{m,n} \mu^z_{m,n,i}(t)\eqperiod{}
\end{equation}
Here, $\mathbb{N} = L_x L_y \rho$ is the total number of molecules in the monolayer and $\mathbb{M}$ is the number of atoms in each molecule.
In writing \refeq{}~\refeqnum{Eq::HallSum_Trial}, we assume a quasistatic approximation can be applied, as was done in the main text.
We denote the mean value of $V_{\text{H},i}(t)$ when $N$ experimental trials are performed as
\begin{equation}\label{Eq::HallSum_Trials}
    \langle V_{\text{H}}(t)\rangle_{N} = \frac{1}{N} \sum^{N}_{i=1} V_{\text{H},i}(t)\eqperiod{}
\end{equation}
The reference experiment averaged measurements from three trials (for each bias point) \cite{2017_Kumar_ChiralityInducedSpinPolarizationPlacesSymmetry}, which corresponds to \refeq{}~\refeqnum{Eq::HallSum_Trials} with $N = 3$. 

The expectation value of $\mu_{m,n,i}^z(t)$ is given by
\begin{equation}\label{Eq::Dipole_Ensemble}
    \langle \mu_{m,n}^z(t) \rangle = \lim_{N \rightarrow \infty} \frac{1}{N} \sum^{N}_{i=1} \mu^z_{m,n,i}(t)\eqcomma{}
\end{equation}
where, unlike the main text, we explicitly use $\langle \, \rangle$ to indicate an expectation value.
By neglecting edge effects, e.g., assuming that on average all molecules in the experiment experience the same bias conditions and interactions, we take $\langle \mu_{m,n}^z(t) \rangle = \langle \mu_{m}^z(t) \rangle$ such that the magnetic dipole moments only depend on the atomic site, not the position of the molecule within the monolayer.
Then, using linearity of the expectation value operator, we write the expectation value of $V_{\text{H},i}(t)$ as
\begin{equation}\label{Eq::HallSum_Expectation}
    \begin{split}
        \langle V_{\text{H}}(t)\rangle &= \sum^{\mathbb{M}}_{m=1} \sum^{\mathbb{N}}_{n=1} w_{m,n} \langle \mu^z_{m,n}(t)\rangle \\
        &= \sum^{\mathbb{M}}_{m=1} \Bigg[\sum^{\mathbb{N}}_{n=1} w_{m,n} \Bigg] \langle \mu^z_{m}(t)\rangle = \sum^{\mathbb{M}}_{m=1} w'_{m} \langle \mu^z_{m}(t)\rangle\eqcomma{}
    \end{split}
\end{equation}
which has an identical form as the equation used to calculate the average magnetic field in the main text. 

Our goal is to show that $\langle V_{\text{H}}(t) \rangle_{N}$, which follows from $N$ experimental trials, can be approximated by the corresponding $\langle V_{\text{H}}(t) \rangle$, which follows from infinitely many experimental trials.
Verification of this approximation would justify our magnetic-field calculation described in the main text, which uses the expectation value of magnetic dipole moments according to \refeq{}~\refeqnum{Eq::HallSum_Expectation}.
Clearly, $\langle V_{\text{H}}(t) \rangle \approx \langle
V_{\text{H}}(t) \rangle_{N}$ is a valid approximation when $N$ is sufficiently large;
the number of trials required to make this approximation depends on the standard deviation of $\langle V_{\text{H}}(t) \rangle_{N}$, denoted $\sigma_{N}$.
Based on error bars in the peak Hall voltage plot of the reference experiment following three trials \cite{2017_Kumar_ChiralityInducedSpinPolarizationPlacesSymmetry}, we estimate $\sigma_{3} / | \langle V_{\text{H}}(t) \rangle_{3} | = 19.4\%$.
% #NOTE: The standard deviation used is truly an estimate.
% The certainty of the standard deviation estimate would improve as more sets of three-trial experiments are performed.
% As it stands, the three monolayers trials, each with 10 non trivial bias points, results in the estimated standard deviation value itself having a standard deviation of 6.2%.
% This is small uncertainty, so our estimated standard deviation can be considered reasonably accurate for our purposes.
Using Chebyshev's inequality \cite{2009_Feller_IntroductionProbabilityTheory}, the condition
\begin{equation}
    | \langle V_{\text{H}}(t) \rangle_{N} - \langle V_{\text{H}}(t) \rangle | \geq k \sigma_{N}
\end{equation}
has at most a $1/k^2$ probability of being satisfied.
Taking $k \sigma_{3} / | \langle V_{\text{H}}(t) \rangle_{3} | = 50\% $, there is at least an $85.0\%$ probability that $\langle V_{\text{H}}(t) \rangle$ is within $50\%$ of $\langle V_{\text{H}}(t) \rangle_{3}$. Thus, $\langle V_{\text{H}}(t) \rangle \approx \langle V_{\text{H}}(t) \rangle_{3}$ to within an order of magnitude.

Ultimately, our magnetic-field calculation described in the main text, which uses the expectation value of magnetic dipole moments according to \refeq{}~\refeqnum{Eq::HallSum_Expectation}, is unlikely to introduce discrepancies exceeding an order of magnitude compared to a corresponding calculation done with \refeq{}~\refeqnum{Eq::HallSum_Trials}, which is sufficient for our purposes.

\section{Magnetic field from uniform surface density of magnetic dipole moments}\label{Sec::DipoleSurfaceMagneticField}

Motivated by discussion in the main text, we determine an analytical expression for the magnetic field arising from a uniform magnetic dipole surface density $\rho \bm{\mu} = \rho [\mu^x,\,\mu^y,\,\mu^z]$ constrained to a rectangular surface given by
\begin{equation}\label{Eq::DipoleSurface}
    S' = \{\bm{r}'=[x',\,y',\,z'] \mid x'\in[0,\,L_x],\,y'\in[0,\,L_y]\}\eqperiod{}
\end{equation}
We limit analysis of the magnetic field to a rectangular surface parallel to the dipole surface but with a $z$ offset $\Delta z = z-z'$.
This surface is given by
\begin{equation}\label{Eq::SensorSurface}
    S = \{\bm{r}=[x,\,y,\,z] \mid x\in[0,\,L_x],\,y\in[0,\,L_y]\}\eqperiod{}
\end{equation}
Primed coordinates indicate the dipole locations, and unprimed coordinates indicate the magnetic-field locations.
Proper selection of $\Delta z$ allows $S'$ to define a sublayer of the monolayer and $S$ to define the Hall sensor surface, which were introduced in the main text.

We use $B^z_{\text{rect}}(x,\,y,\,\Delta z)$ to denote the magnetic field over $S$ from a dipole surface density constrained to $S'$.
The average of $B^z_{\text{rect}}(x,\,y,\,\Delta z)$ over $S$ is denoted $\bar{B}^z_{\text{rect}}(\Delta z)$.
The notation $\bar{B}$ is used throughout this analysis to indicate a spatially averaged magnetic field over $S$.
An expression for $\bar{B}^z_{\text{rect}}(\Delta z)$ is our end goal because, as justified in Section~\ref{Sec::HallSensorWeighting}, we assume $B_{\text{rect}}^z(x,\,y,\,\Delta z)$ provides a contribution to a Hall voltage measurement that is proportional to $\bar{B}_{\text{rect}}^z(\Delta z)$.

To start, the magnetic field from a single magnetic dipole $\bm{\mu}$ located at $\bm{r}'$ is 
\begin{equation}\label{Eq::DipoleMagneticField}
    \bm{B}_{\text{dipole}}(\bm{r},\,\bm{r}') = \frac{\mu_0}{4\pi} \bigg\{ \frac{3 (\bm{r}-\bm{r}')[\bm{\mu}\cdot(\bm{r}-\bm{r}')]}{|\bm{r}-\bm{r}'|^5} - \frac{\bm{\mu}}{|\bm{r}-\bm{r}'|^3}\bigg\}\eqperiod{}
\end{equation}
The $\hat{\bm{z}}$ component of the magnetic field is the only component relevant to the Hall sensor measurement, so we consider
\begin{equation}\label{Eq::DipoleMagneticField_ZComponent}
    \begin{split}
        B_{\text{dipole}}^z(\bm{r},\,\bm{r}') ={}& \frac{\mu_0\mu^x}{4\pi} \, \frac{3(x-x')(z-z')}{|\bm{r}-\bm{r}'|^5} \\
        &+ \frac{\mu_0\mu^y}{4\pi} \, \frac{3(y-y')(z-z')}{|\bm{r}-\bm{r}'|^5}\\
        &+ \frac{\mu_0\mu^z}{4\pi} \bigg[ \frac{3 (z-z')^2}{|\bm{r}-\bm{r}'|^5} - \frac{1}{|\bm{r}-\bm{r}'|^3}\bigg]\eqcomma{}
    \end{split}
\end{equation}
which is still dependent on each component of $\bm{\mu}$.

To find $B_{\text{rect}}^z(x,\,y,\,\Delta z)$, we replace $\bm{\mu}$ in \refeq{}~\refeqnum{Eq::DipoleMagneticField_ZComponent} with an infinitesimal dipole surface element $\rho \bm{\mu} \, \text{d}x' \, \text{d}y'$ then integrate over $S'$ such that
\begin{equation}\label{Eq::DipoleSurfaceField} 
    \begin{split}
        B_{\text{rect}}^z(x,\,y,\,\Delta z) 
        ={}& B_{\text{rect},x}^z(x,\,y,\,\Delta z) + B_{\text{rect},y}^z(x,\,y,\,\Delta z) \\
        &+ B_{\text{rect},z}^z(x,\,y,\,\Delta z)\eqcomma{}
    \end{split}
\end{equation}
where
\begin{equation}\label{Eq::XDipoleSurface} 
    B_{\text{rect},x}^z(x,\,y,\,\Delta z) = \frac{\mu_0\rho\mu^x}{4\pi} \int_0^{L_x} \text{d}x' \int_0^{L_y} \text{d}y' \, \frac{3(x-x')\Delta z}{|\bm{r}-\bm{r}'|^5}\eqcomma{}
\end{equation}
\begin{equation}\label{Eq::YDipoleSurface} 
    B_{\text{rect},y}^z(x,\,y,\,\Delta z) =  \frac{\mu_0\rho\mu^y}{4\pi} \int_0^{L_x} \text{d}x' \int_0^{L_y} \text{d}y' \, \frac{3(y-y')\Delta z}{|\bm{r}-\bm{r}'|^5}\eqcomma{}
\end{equation}
and
\begin{equation}\label{Eq::ZDipoleSurface}
    \begin{split}
        B_{\text{rect},z}^z(x,\,y,\,\Delta z) ={}& \frac{\mu_0\rho\mu^z}{4\pi} \int_0^{L_x} \text{d}x' \int_0^{L_y} \text{d}y' \, \\
        &\times \bigg[ \frac{3 \Delta z^2}{|\bm{r}-\bm{r}'|^5} - \frac{1}{|\bm{r}-\bm{r}'|^3}\bigg] \eqperiod{}
    \end{split}
\end{equation}

\begin{figure}[htbp!]
    \captionsetup[subfloat]{farskip=0pt,captionskip=0pt}
    \subfloat[][]{\includegraphics{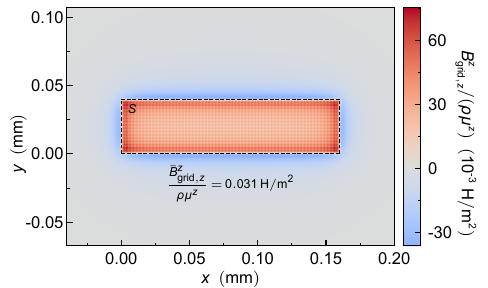}\label{Fig::MagneticField_Numerical_Sparse}}\\
    \subfloat[][]{\includegraphics{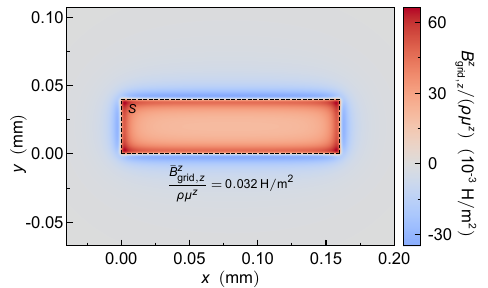}\label{Fig::MagneticField_Numerical_Dense}}
    \caption{
        Magnetic field from a discrete grid of magnetic dipoles with magnetic moment $\bm{\mu} = \mu^z \hat{\bm{z}}$, where dipoles are separated by (a) $1/\sqrt{\rho} = 2.4 \; \mathrm{\upmu m}$ and (b) $1/\sqrt{\rho} =  0.1 \; \mathrm{\upmu m}$.
        The sensor surface $S$ and dipole grid are separated by $\Delta z = 2.5 \; \mathrm{\upmu m}$, and both have dimensions $L_x = 160 \; \mathrm{\upmu m}$ and $L_y = 40 \; \mathrm{\upmu m}$.
        Results were calculated using discrete Fourier analysis and are normalized by $\rho \mu^z$.
    }
    \label{Fig::MagneticField_Numerical}
\end{figure}

As an aside: Each sublayer of the monolayer described in the main text is more precisely defined as a discrete grid of dipoles as opposed to a dipole surface density of dipoles.
However, in the limit $\rho L_x L_y \rightarrow \infty$, the magnetic field arising from a discrete grid of dipoles constrained to $S'$, denoted $B_{\text{grid}}^z(x,\,y,\,\Delta z)$, is equivalent to $B^z_{\text{rect}}(x,\,y,\,\Delta z)$.
This equivalence follows from converting the sum $B^z_{\text{grid}}(x,\,y,\,\Delta z) = \sum_{x',y'} B_{\text{dipole}}^z(\bm{r},\,\bm{r}')$ into an integral.
In particular, we write
\begin{equation}\label{Eq::GridRectEquivalence}
    \bar{B}_{\text{grid}}^z(\Delta z)\approx \bar{B}_{\text{rect}}^z(\Delta z),\ \text{if } \rho L_x L_y \gg 1\eqcomma{}
\end{equation}
which we will verify later. Using parameters from the main text, we estimate that each sublayer used to model the reference experiment contains $\rho L_x L_y \approx 1.7\times 10^{10}$ dipoles, which is sufficiently large for the purposes of \refeq{}~\refeqnum{Eq::GridRectEquivalence}. Naively  summing \refeq{}~\refeqnum{Eq::DipoleMagneticField_ZComponent} to calculate $\bar{B}^z_{\text{grid},z}(\Delta z)$ becomes infeasible for increasing quantities of discrete dipoles.
Discrete Fourier analysis alleviates the dependence of the computational cost on $\rho$, but we notice that the number of numerical grid points required for the convergence of $\bar{B}^z_{\text{grid},z}(\Delta z)$ increases with $L_x L_y/\Delta z^2$.
% #NOTE: It seems that the separation between numerical grid points should be roughly half of $\Delta_z$ for reasonable level of convergence.
When $L_x L_y/\Delta z^2$ is small such that discrete Fourier analysis is feasible, $B_{\text{grid}}^z(x,\,y,\,\Delta z)$ is shown in \reffig{}~\ref{Fig::MagneticField_Numerical} for two different $\rho$ values.
% #NOTE: \Reffig{}~\ref{Fig::MagneticField_Numerical} indicates that $\bar{B}^z_{\text{grid},z}(\Delta z)/\rho$ is not strongly influenced by changes to $\rho$ if $\rho L_x L_y \gg 1$.
At the dimensions required to model the reference experiment, discrete Fourier analysis is inefficient, which motivates the use of a dipole surface density according to \refeq{}~\refeqnum{Eq::DipoleSurfaceField} in the first place.

\begin{figure}[htbp!]
    \captionsetup[subfloat]{farskip=0pt,captionskip=0pt}
    \subfloat[][]{\includegraphics{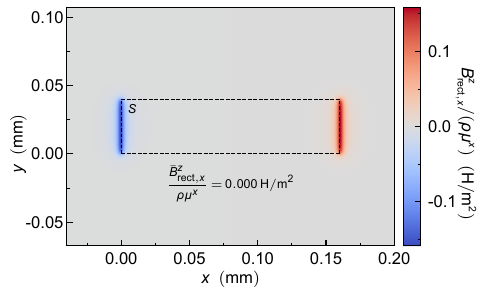}\label{Fig::MagneticField_XDipole}}\\
    \subfloat[][]{\includegraphics{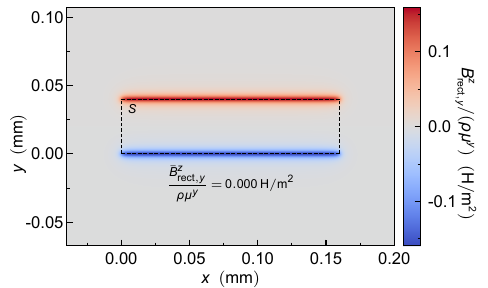}\label{Fig::MagneticField_YDipole}}
    \caption{
        Magnetic field from a rectangular surface density of magnetic dipoles with magnetic moment (a) $\bm{\mu} = \mu^x \hat{\bm{x}}$ and (b) $\bm{\mu} = \mu^y \hat{\bm{y}}$.
        The sensor surface $S$ and dipole surface are separated by $\Delta z = 2.5 \; \mathrm{\upmu m}$, and both have dimensions $L_x = 160 \; \mathrm{\upmu m}$ and $L_y = 40 \; \mathrm{\upmu m}$.
        Results were calculated according to \refeqs{}~\refeqnum{Eq::XDipoleSurface_Analytic} and \refeqnum{Eq::YDipoleSurface_Analytic} for (a) and (b), respectively, and are normalized by $\rho |\bm{\mu}|$.
    }
\end{figure}

Analytical solutions to \refeqs{}~\refeqnum{Eq::XDipoleSurface} and \refeqnum{Eq::YDipoleSurface} are straightforward to determine.
Focusing on \refeqs{}~\refeqnum{Eq::XDipoleSurface}, the substitutions $(x-x')^2 = u$ and $(y-y') = \sqrt{ (x-x')^2 + \Delta z^2} \tan{\theta}$ lead to
\begin{equation}\label{Eq::XDipoleSurface_Analytic}
    \begin{split}
        B_{\text{rect},x}^z(x,\,y,\,\Delta z) ={}& -\frac{\mu_0\rho\mu^x}{4\pi} [F(x-L_x,\, y-L_y,\, \Delta z) \\
        &- F(x-L_x,\, y,\, \Delta z) - F(x,\, y-L_y,\, \Delta z) \\
        &+ F(x,\, y,\, \Delta z)]\eqcomma{}
    \end{split}
\end{equation}
where 
\begin{equation}
    F(\alpha,\,\beta,\,\Delta z) = \frac{\beta \Delta z}{(\alpha^2+\Delta z^2) \sqrt{\alpha^2 + \beta^2 + \Delta z^2}}\eqperiod{}
\end{equation}
Because \refeq{}~\refeqnum{Eq::YDipoleSurface} has the same form as \refeq{}~\refeqnum{Eq::XDipoleSurface}, we have
\begin{equation}\label{Eq::YDipoleSurface_Analytic}
    \begin{split}
        B_{\text{rect},y}^z(x,\,y,\,\Delta z) ={}& -\frac{\mu_0\rho\mu^y}{4\pi} [ F(y-L_y,\, x-L_x,\, \Delta z) \\
        &- F(y-L_y,\, x,\, \Delta z)- F(y,\, x-L_x,\, \Delta z) \\
        &+ F(y,\, x,\, \Delta z) ]\eqperiod{}
    \end{split}
\end{equation}
Plots of $B_{\text{rect},x}^z(x,\,y,\,\Delta z)$ and $B_{\text{rect},y}^z(x,\,y,\,\Delta z)$ are shown in \reffigs{}~\ref{Fig::MagneticField_XDipole} and \ref{Fig::MagneticField_YDipole}, respectively.
Clearly, from symmetry, $\bar{B}_{\text{rect},x}^z(\Delta z) = \bar{B}_{\text{rect},y}^z(\Delta z) = 0$.
Only the edges of $S$ are responsible for the magnetic fields, which will be discussed in more detail during the derivation of $B_{\text{rect},z}^z(x,\,y,\,\Delta z)$.

Finding an exact analytical solution to \refeq{}~\refeqnum{Eq::ZDipoleSurface} is less straightforward, so we present two approaches: (1) We reduce the four-fold integral required to determine $\bar{B}^z_{\text{rect},z}(\Delta z)$ to a two-fold integral, which provides a computationally inexpensive  numerical solution to $\bar{B}^z_{\text{rect},z}(\Delta z)$. (2) We employ approximations to obtain analytical expressions for both $B^z_{\text{rect},z}(x,\,y,\,\Delta z)$ and $\bar{B}^z_{\text{rect},z}(\Delta z)$. Unlike the first approach, the second approach returns an expression for $B^z_{\text{rect},z}(x,\,y,\,\Delta z)$, which will be necessary when considering the effects of Hall sensor nonuniformity in Section~\ref{Sec::HallSensorWeighting}.

For the first approach, integrating \refeq{}~\refeqnum{Eq::ZDipoleSurface} over $x \in [0,\,L_x]$ and $y \in [0,\,L_y]$ and dividing by $L_x L_y$ results in an equation with the general form
\begin{equation}
    \begin{split}
        \bar{B}_{\text{rect}}(\Delta z) ={}& \frac{1}{L_x L_y} \int^{L_x}_0 \text{d}x \int^{L_y}_0 \text{d}y \int^{L_x}_0 \text{d}x' \int^{L_y}_0 \text{d}y' \, \\
        &\times B_{\text{dipole}}(x-x',\, y-y',\,\Delta z)\eqperiod{}
    \end{split}
\end{equation}
Setting $x-x' = \tilde{x}$, $x+x' = \bar{x}$,  $y-y' = \tilde{y}$, and $y+y' = \bar{y}$ allows for $\text{d}x\, \text{d}x' = -\frac{1}{2} \text{d}\tilde{x}\, \text{d}\bar{x}$ and $\text{d}y\, \text{d}y' = -\frac{1}{2} \text{d}\tilde{y}\, \text{d}\bar{y}$. Integrating over slices of $\tilde{x}$ and $\tilde{y}$ gives
\begin{equation}
    \begin{split}
        \bar{B}(\Delta z) ={}& \frac{1}{4 L_x L_y} \\
        &\times \bigg[\int^{L_x}_0 \text{d}\tilde{x} \int^{-\tilde{x} + 2 L_x}_{\tilde{x}} \text{d}\bar{x} + \int^{0}_{-L_x} \text{d}\tilde{x} \int^{\tilde{x} + 2 L_x}_{-\tilde{x}} \text{d}\bar{x} \bigg] \\
        &\times \bigg[\int^{L_y}_0 \text{d}\tilde{y} \int^{-\tilde{y} + 2 L_y}_{\tilde{y}} \text{d}\bar{y} + \int^{0}_{-L_y} \text{d}\tilde{y} \int^{\tilde{y} + 2 L_y}_{-\tilde{y}} \text{d}\bar{y} \bigg] \\
        &\times B_{\text{dipole}}(\tilde{x},\, \tilde{y},\, \Delta z)\eqperiod{}
    \end{split}
\end{equation}
The integrand is independent of $\bar{x}$ and $\bar{y}$, so evaluating gives
\begin{equation}
    \begin{split}
        \bar{B}_{\text{rect}}(\Delta z) ={}& \frac{1}{L_x L_y} \\
        &\times \bigg[\int^{L_x}_0 \text{d}\tilde{x} \, (L_x - \tilde{x}) - \int^{-L_x}_{0} \text{d}\tilde{x} \, (L_x + \tilde{x}) \bigg] \\
        &\times \bigg[\int^{L_y}_0 \text{d}\tilde{y} \, (L_y - \tilde{y}) - \int^{-L_y}_{0} \text{d}\tilde{y} \, (L_y + \tilde{y}) \bigg] \\
        &\times B_{\text{dipole}}(\tilde{x},\, \tilde{y},\, \Delta z)\eqperiod{}
    \end{split}
\end{equation}
Substituting $\tilde{x}$ and $\tilde{y}$ with their negative counterparts in the second and fourth integrals allows combining integrals with different integrands
\begin{equation}\label{Eq::GeneralDipoleSurface_Average}
    \begin{split}
        \bar{B}_{\text{rect}}(\Delta z) ={}& \frac{1}{L_x L_y} \int^{L_x}_0 \text{d}\tilde{x} \, (L_x - \tilde{x})  \int^{L_y}_0 \text{d}\tilde{y} \, (L_y - \tilde{y}) \\
        &\times [ B_{\text{dipole}}(\tilde{x},\, \tilde{y},\, \Delta z) + B_{\text{dipole}}(-\tilde{x},\, \tilde{y},\, \Delta z) \\
        &+ B_{\text{dipole}}(\tilde{x},\, -\tilde{y},\, \Delta z) + B_{\text{dipole}}(-\tilde{x},\, -\tilde{y},\, \Delta z)]\eqcomma{}
    \end{split}
\end{equation}
which, in the case of \refeq{}~\refeqnum{Eq::ZDipoleSurface}, simplifies to
\begin{equation}\label{Eq::ZDipoleSurface_Average}
    \begin{split}
        \bar{B}^z_{\text{rect},z}(\Delta z) ={}& \frac{4}{L_x L_y} \int^{L_x}_0 \text{d}\tilde{x} \, (L_x - \tilde{x})  \int^{L_y}_0 \text{d}\tilde{y} \, (L_y - \tilde{y}) \\
        &\times B^z_{\text{dipole},z}(\tilde{x},\, \tilde{y},\, \Delta z)\eqcomma{}
    \end{split}
\end{equation}
where $B^z_{\text{dipole},z}(x-x',\, y-y',\, \Delta z)$ is the third term of \refeq{}~\refeqnum{Eq::DipoleMagneticField_ZComponent}. Thus, we have reduced the four-fold integral to a two-fold integral that can be evaluated numerically.
% #NOTE: Further numerical benefits can be gained by converting to polar coordinates in $\tilde{x}$ and $\tilde{y}$.

\begin{figure}[htbp!]
    \captionsetup[subfloat]{farskip=0pt,captionskip=0pt}
    \subfloat[][]{\includegraphics{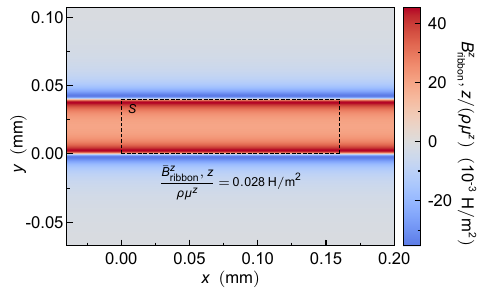}\label{Fig::MagneticField_LowerBound}}\\
    \subfloat[][]{\includegraphics{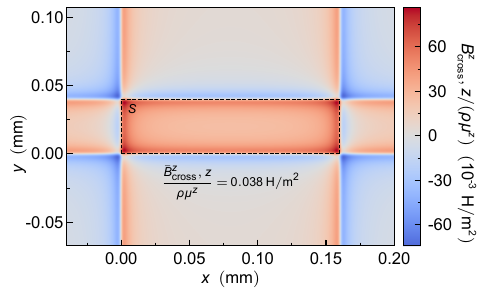}\label{Fig::MagneticField_UpperBound}}
    \caption{
        Magnetic field from a surface density of magnetic dipoles with magnetic moment $\bm{\mu} = \mu^z \hat{\bm{z}}$, where the dipole surface is a (a) ribbon and (b) cross.
        The sensor surface $S$ has dimensions $L_x = 160 \; \mathrm{\upmu m}$ and $L_y = 40 \; \mathrm{\upmu m}$ and is separated from the dipole surface by $\Delta z = 2.5 \; \mathrm{\upmu m}$.
        Results were calculated according to \refeqs{}~\refeqnum{Eq::ZDipoleSurface_Ribbon} and \refeqnum{Eq::ZDipoleSurface_Cross} for (a) and (b), respectively, and are normalized by $\rho \mu^z$.
    }
\end{figure}

For the second approach, to find an analytical approximation for $B^z_{\text{rect},z}(x,\,y,\,\Delta z)$ and $\bar{B}^z_{\text{rect},z}(\Delta z)$, we use the fact that $S'$ is a long, thin, rectangle with $L_x = 500\;\mathrm{\upmu m}$ and $L_y = 40\;\mathrm{\upmu m}$, based on dimensions from the reference experiment \cite{2017_Kumar_ChiralityInducedSpinPolarizationPlacesSymmetry}.
Taking the $L_x/L_y$ ratio to the extreme, we consider a dipole surface defined by
\begin{equation}
    S'_{\text{ribbon}} = \{\bm{r}'=[x',\,y',\,z'] \mid x' \in (-\infty,\, +\infty),\,y'\in[0,\,L_y]\}\eqcomma{}
\end{equation}
which is a ribbon with width $L_y$ and infinite length. The resulting magnetic field is
\begin{equation}
    \begin{split}
        B_{\text{ribbon},z}^z(x,\,y,\,\Delta z) ={}& \frac{\mu_0\rho\mu^z}{4\pi} \int_{-\infty}^{\infty} \text{d}x' \int_0^{L_y} \text{d}y' \, \\
        & \times \bigg[\frac{3 \Delta z^2}{|\bm{r}-\bm{r}'|^5} - \frac{1}{|\bm{r}-\bm{r}'|^3}\bigg]\eqperiod{}
    \end{split}
\end{equation}
Substituting $(x-x') = \sqrt{(y-y')^2 + \Delta z^2} \tan{\theta}$ and $(y-y') = \Delta z\tan{\theta'}$ leads to 
\begin{equation}\label{Eq::ZDipoleSurface_Ribbon}
    B_{\text{ribbon},z}^z(x,\,y,\,\Delta z) = 2 \frac{\mu_0\rho\mu^z}{4\pi} \bigg[ \frac{y}{y^2+\Delta z^2} - \frac{y-L_y}{(y-L_y)^2+\Delta z^2} \bigg]\eqcomma{}
\end{equation}
which reduces to the equation for a semi-infinite dipole surface \cite{2009_Votyakov_AnalyticModelsHeterogenousMagnetic} in the limit $L_y \rightarrow \infty$.
A plot of $B_{\text{ribbon},z}^z(x,\,y,\,\Delta z)$ is shown in \reffig{}~\ref{Fig::MagneticField_LowerBound}, where we mark the finite extent of $S$.
The magnetic field is independent of $x$, which is expected because a ribbon with infinite $x$ extent has translational symmetry along $x$.
Edge effects are visible along $y=0$ and $y=L_y$, but because the dipole surface extends to infinity in $x$, no such edge effects are seen along $x=0$ and $x=L_x$. For this reason, $B^z_{\text{ribbon},z}(x,\,y,\,\Delta z)$ includes less edge effects over $S$ compared to $B^z_{\text{rect},z}(x,\,y,\,\Delta z)$.

To understand how $\bar{B}^z_{\text{ribbon},z}(\Delta z)$ compares to $\bar{B}^z_{\text{rect},z}(\Delta z)$, it is instructive to consider the magnetic field resulting from an infinite dipole surface.
Translating \refeq{}~\refeqnum{Eq::ZDipoleSurface_Ribbon} along $y$ by $-L_y/2$ and taking the limit $L_y \rightarrow \infty$ gives
\begin{equation}\label{Eq::ZDipoleSurface_Infinite}
    \begin{split}
        B_{\text{infinite},z}^z(x,\,y,\,\Delta z) ={}& \lim_{L_y\rightarrow\infty} 2 \frac{\mu_0\rho\mu^z}{4\pi} \bigg[ \frac{y+L_y/2}{(y+L_y/2)^2+\Delta z^2} \\
        &- \frac{y-L_y/2}{(y-L_y/2)^2+\Delta z^2} \bigg] \\
        ={}& 2 \frac{\mu_0\rho\mu^z}{4\pi} (0 + 0) = 0\eqperiod{}
    \end{split}
\end{equation}
Hence, an infinite dipole surface does not generate a magnetic field.
This result also follows directly from Maxwell's equations, which require that the normal component of $\bm{B}$ must be continuous across any interface.
\Refeq{}~\refeqnum{Eq::ZDipoleSurface_Infinite} indicates that edge effects are responsible for the magnetic field of a uniform dipole surface, which is consistent with our results in \reffigs{}~\ref{Fig::MagneticField_XDipole} and \ref{Fig::MagneticField_YDipole}.
Then, we expect that $|\bar{B}^z_{\text{ribbon},z}(\Delta z)|$ is a lower bound for $|\bar{B}^z_{\text{rect},z}(\Delta z)|$, which we verify later.

We now aim to find an upper bound for $|\bar{B}^z_{\text{rect},z}(\Delta z)|$, which would constrain $\bar{B}^z_{\text{rect},z}(\Delta z)$ to a known range.
Based on \refeq{}~\refeqnum{Eq::ZDipoleSurface_Ribbon}, the magnetic field from another dipole ribbon, which has a width $L_x$ and infinite $y$ extent, is given by 
\begin{equation}\label{Eq::ZDipoleSurface_Ribbon2}
    B_{\text{ribbon'},z}^z(x,\,y,\,\Delta z) = \frac{\mu_0\rho\mu^z}{4\pi} \bigg[ \frac{x}{x^2+\Delta z^2} - \frac{x-L_x}{(x-L_x)^2+\Delta z^2} \bigg]\eqperiod{}
\end{equation}
This magnetic field only includes edge effects along $x=0$ and $x=L_x$.
Superimposing the dipole ribbons corresponding to \refeq{}~\refeqnum{Eq::ZDipoleSurface_Ribbon} and \refeqnum{Eq::ZDipoleSurface_Ribbon2} with an infinite dipole surface of opposite sign $-\rho \mu^z$ leads to a dipole surface with density $\rho \mu^z$ over $S'$, but dipole surfaces with density $-\rho \mu^z$ extending to infinity from each corner of $S'$.
The magnetic field resulting from the superimposition of these dipole surfaces is given by
\begin{equation}\label{Eq::ZDipoleSurface_Cross}
    \begin{split}
        B_{\text{cross},z}^z(x,\,y,\,\Delta z) ={}& B_{\text{ribbon},z}^z(x,\,y,\,\Delta z) + B_{\text{ribbon'},z}^z(x,\,y,\,\Delta z) \\
        &- B_{\text{infinite},z}^z(x,\,y,\,\Delta z) \\
        ={}& B_{\text{ribbon},z}^z(x,\,y,\,\Delta z) + B_{\text{ribbon'},z}^z(x,\,y,\,\Delta z) \\
        &- 0\eqperiod{}
    \end{split}
\end{equation}
A plot of $B_{\text{cross},z}^z(x,\,y,\,\Delta z)$ is shown in \reffig{}~\ref{Fig::MagneticField_UpperBound}, where we mark the finite extent of the sensor surface $S$.
The inclusion of both dipole ribbons causes $B_{\text{cross},z}^z(x,\,y,\,\Delta z)$ to overestimate edge effects at the corners of $S$ compared to $B^z_{\text{rect},z}(x,\,y,\,\Delta z)$.
Then, we expect that $|\bar{B}^z_{\text{cross},z}(\Delta z)|$ is an upper bound for $|\bar{B}^z_{\text{rect},z}(\Delta z)|$, which we verify later.

To determine $\bar{B}^z_{\text{ribbon},z}(\Delta z)$ and $\bar{B}^z_{\text{cross},z}(\Delta z)$, we start by integrating \refeq{}~\refeqnum{Eq::ZDipoleSurface_Ribbon} according to
\begin{equation}\label{Eq::ZDipoleSurface_Ribbon_Average}
    \begin{split}
        \bar{B}_{\text{ribbon},z}^z(\Delta z) ={}& \frac{2}{L_x L_y} \, \frac{\mu_0\rho\mu^z}{4\pi} \int_0^{L_x} \text{d}x \int_0^{L_y} \text{d}y \, \\
        &\times \bigg[ \frac{y}{y^2+\Delta z^2} - \frac{y-L_y}{(y-L_y)^2+\Delta z^2} \bigg]\eqperiod{}
    \end{split}
\end{equation}
Substituting $y^2=u$ and $(y-L_y)^2=u$ for the first and second terms of \refeq{}~~\refeqnum{Eq::ZDipoleSurface_Ribbon_Average}, respectively, leads to
\begin{equation}
    \bar{B}_{\text{ribbon},z}^z(\Delta z) = 2 \frac{\mu_0\rho\mu^z}{4\pi} \, \frac{1}{L_y} \ln{\bigg(\frac{{L_y}^2}{\Delta z^2} + 1\bigg)}\eqperiod{}
\end{equation}
Similarly, $\bar{B}^z_{\text{cross},z}(\Delta z)$ follows as 
\begin{equation}
    \begin{split}
        \bar{B}_{\text{cross},z}^z(\Delta z) ={}& 2 \frac{\mu_0\rho\mu^z}{4\pi} \bigg[ \frac{1}{L_y} \ln{\bigg(\frac{{L_y}^2}{\Delta z^2} + 1\bigg)} \\
        &+ \frac{1}{L_x} \ln{\bigg(\frac{{L_x}^2}{\Delta z^2} + 1\bigg)} \bigg]\eqperiod{}
    \end{split}
\end{equation}

\begin{figure}[htbp!]
    \captionsetup[subfloat]{farskip=0pt,captionskip=0pt}
    \subfloat[][]{\includegraphics{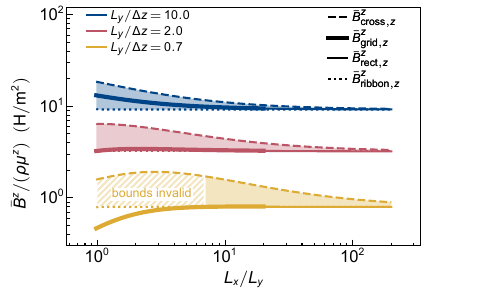}\label{Fig::MagneticField_BoundingDimensionSweep}}\\
    \subfloat[][]{\includegraphics{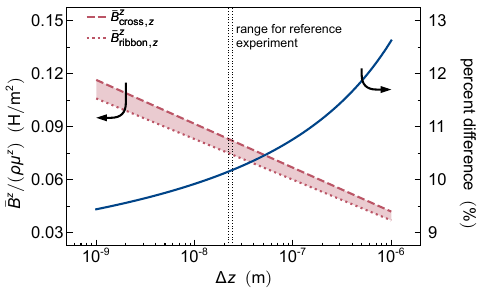}\label{Fig::MagneticField_BoundingUncertainty}}
    \caption{
        (a) Comparison of $\bar{B}^z_{\text{rect},z}(\Delta z)$, $\bar{B}^z_{\text{ribbon},z}(\Delta z)$, $\bar{B}^z_{\text{cross},z}(\Delta z)$, and $\bar{B}^z_{\text{grid},z}(\Delta z)$ as a function of $L_x/L_y$ for various $L_y/\Delta z$, with $L_y = 100 \; \text{nm}$ fixed.
        The separation between dipoles for the discrete grid is $1/\sqrt{\rho} = 0.1 \; \text{nm}$.
        (b) Comparison between $\bar{B}^z_{\text{ribbon},z}(\Delta z)$ and $\bar{B}^z_{\text{cross},z}(\Delta z)$, including  percent difference, as a function of $\Delta z$, with $L_x = 500 \; \mathrm{\upmu m}$ and $L_y = 40 \; \mathrm{\upmu m}$.
    }
\end{figure}

Plots of $\bar{B}^z_{\text{rect},z}(\Delta z)$, $\bar{B}^z_{\text{ribbon},z}(\Delta z)$, $\bar{B}^z_{\text{cross},z}(\Delta z)$, and $\bar{B}^z_{\text{grid},z}(\Delta z)$ as a function of $L_x/L_y$ for various $L_y/\Delta z$ are shown in \reffig{}~\ref{Fig::MagneticField_BoundingDimensionSweep}, which illustrates that
\begin{equation}\label{Eq::MagneticField_BoundsCondidtions}
    \begin{split}
        |\bar{B}^z_{\text{ribbon},z}(\Delta z)| < |\bar{B}^z_{\text{rect},z}(\Delta z)| < |\bar{B}^z_{\text{cross},z}(\Delta z)|,\\
        \text{if  } 2\Delta z \alt L_y \alt L_x \eqperiod{}
    \end{split}
\end{equation}
% #NOTE: The value of 2 in the inequality is directly suggested by the associated Figure, but I have also checked this more intentionally with a smaller numerical grid, and it seems to hold.
% Because I have no analytical proof, I use approximate inequalities. 
% In the main text, I simply use much less then sign, which is more conservative.
We have verified that \refeq{}~\refeqnum{Eq::MagneticField_BoundsCondidtions} holds for $L_x L_y/\Delta z^2$ as large as $L_x/\Delta z = L_y/\Delta z = 5\times10^4$.
% #NOTE: Tested up to $L_x/\Delta z = L_y/\Delta z = 50000$.
Further, the agreement between $\bar{B}^z_{\text{rect},z}(\Delta z)$ and $\bar{B}^z_{\text{grid},z}(\Delta z)$ supports the validity of \refeq{}~\refeqnum{Eq::GridRectEquivalence}.
Based on \refeq{}~\refeqnum{Eq::MagneticField_BoundsCondidtions}, we can approximate \refeq{}~\refeqnum{Eq::ZDipoleSurface_Average} according to 
\begin{equation}\label{Eq::ZDipoleSurface_Approximation_Average}
    \begin{split}
        \bar{B}^z_{\text{rect},z}(\Delta z) \approx{}& \frac{\bar{B}^z_{\text{ribbon},z}(\Delta z) + \bar{B}^z_{\text{cross},z}(\Delta z)}{2}
        = 2 \frac{\mu_0\rho\mu^z}{4\pi} \\
        &\times \bigg[ \frac{1}{L_y} \ln{\bigg(\frac{{L_y}^2}{\Delta z^2} + 1\bigg)} + \frac{1}{2L_x} \ln{\bigg(\frac{{L_x}^2}{\Delta z^2} + 1\bigg)} \bigg]\eqcomma{}
    \end{split}
\end{equation}
which has a corresponding magnetic field distribution given by
\begin{equation}\label{Eq::ZDipoleSurface_Approximation}
    \begin{split}
        B^z_{\text{rect},z}(x,\,y,\,\Delta z) \approx{}& \frac{B^z_{\text{ribbon},z}(x,\,y,\,\Delta z) + B^z_{\text{cross},z}(x,\,y,\,\Delta z)}{2} \\
        ={}& 2 \frac{\mu_0\rho\mu^z}{4\pi} \bigg[F(L_y-y,\, \Delta z) + F_m(y,\, \Delta z) \\
        &+ \frac{F(L_x-x,\, \Delta z)}{2} + \frac{F(x,\, \Delta z)}{2}\bigg]\eqcomma{}
    \end{split}
\end{equation}
where
\begin{equation}
    F_m(\alpha) = \frac{\alpha \Delta z}{\alpha^2 + \Delta z^2}\eqperiod{}
\end{equation}
We emphasize that \refeq{}~\refeqnum{Eq::ZDipoleSurface_Approximation} is only reasonable within $S$ and does not attempt to accurately model the magnetic field outside of $S$.

Taking $L_x = 500\;\mathrm{\upmu m}$ and $L_y = 40\;\mathrm{\upmu m}$ based on dimensions from the reference experiment, we calculate the percent difference between $\bar{B}^z_{\text{ribbon},z}(\Delta z)$ and $\bar{B}^z_{\text{cross},z}(\Delta z)$ for various $\Delta z$.
The result is shown in \reffig{}~\ref{Fig::MagneticField_BoundingUncertainty}, which illustrates that the largest difference between $\bar{B}^z_{\text{ribbon},z}(\Delta z)$ and $\bar{B}^z_{\text{cross},z}(\Delta z)$ comes for large $\Delta z$.
The maximum difference at $\Delta z$ relevant to the reference experiment is 10.2\%.
Then, based on \refeqs{}~\refeqnum{Eq::MagneticField_BoundsCondidtions} and \refeqnum{Eq::GridRectEquivalence}, the corresponding maximum error between $\bar{B}^z_{\text{grid},z}(\Delta z)$ and \refeq{}~\refeqnum{Eq::ZDipoleSurface_Approximation_Average} is 5.36\%.

Ultimately, because $\bar{B}^z_{\text{rect},x}(\Delta z) = 0$ and $\bar{B}^z_{\text{rect},y}(\Delta z) = 0$, we have 
\begin{equation}
    \begin{split}
        \bar{B}^z_{\text{rect}}(\Delta z) &= \bar{B}^z_{\text{rect},x}(\Delta z) + \bar{B}^z_{\text{rect},y}(\Delta z) + \bar{B}^z_{\text{rect},z}(\Delta z) \\
        & \approx \frac{\bar{B}^z_{\text{ribbon},z}(\Delta z) + \bar{B}^z_{\text{cross},z}(\Delta z)}{2}\eqcomma{}
    \end{split}
\end{equation}
which is the result used in the main text.

\section{Hall sensor voltage from weighted average of magnetic field}\label{Sec::HallSensorWeighting}

The Hall voltage measured across the $y$ dimension of a Hall sensor in the presence of a uniform $\hat{\bm{z}}$-directed magnetic field $B^z_{\text{H}}$ is given by \cite{1994_Mantorov_InterpretationHallEffectMeasurements}
\begin{equation}\label{Eq::HallVoltage_Uniform1}
    V_{\text{H}} = \frac{R_{\text{H}} I^x B^z_{\text{H}}}{L_z}\eqcomma{}
\end{equation}
where $R_{\text{H}}$ is the Hall constant, $I^x$ is an $\hat{\bm{x}}$-directed current flowing through the Hall sensor, and $L_z$ is the $z$-thickness of the Hall sensor.
Equivalently, we can write \refeq{}~\refeqnum{Eq::HallVoltage_Uniform1} in terms of a current density $j^x = I^x / (L_y L_z)$ according to
\begin{equation}\label{Eq::HallVoltage_Uniform}
    V_{\text{H}} = R_{\text{H}} j^x L_y B^z_{\text{H}}\eqcomma{}
\end{equation}
where $L_y$ is the $y$ dimension of the Hall sensor.

\Refeq{}~\refeqnum{Eq::HallVoltage_Uniform1} does not apply for nonuniform magnetic fields because, in general, the sensitivity of a Hall sensor has a positional dependence \cite{1994_Mantorov_InterpretationHallEffectMeasurements}.
That is, magnetic fields at different points throughout the Hall sensor may contribute disproportionately to the Hall voltage.
The magnetic fields presented in Section~\ref{Sec::DipoleSurfaceMagneticField} are nonuniform, so the impact, if any, of such nonuniformity on the Hall voltage should be considered. 
Analogous to \refeq{}~\refeqnum{Eq::HallVoltage_Uniform}, we can write the Hall voltage arising from a nonuniform magnetic field as
\begin{equation}\label{Eq::HallVoltage_Nonuniform}
    V_{\text{H}} = R_{\text{H}} j^x L_y \bar{B}^z_{\text{H,eff}}\eqcomma{}
\end{equation}
where 
\begin{equation}\label{Eq::EffectiveMagneticField}
    \bar{B}^z_{\text{H,eff}} = \frac{1}{{L_x L_y}} \int_{0}^{L_x} \text{d}x\, \int_{0}^{L_y} \text{d}y\, B^z_{\text{H}}(x,\,y) S_{\text{H}}(x,\,y)
\end{equation}
is the effective magnetic field measured by the Hall sensor.
\Refeq{}~\refeqnum{Eq::EffectiveMagneticField} constitutes a weighted spatial average of the magnetic field $B^z_{\text{H}}(x,\,y)$ over the Hall sensor surface based on the Hall sensor sensitivity $S_{\text{H}}(x,\,y)$.
In contrast, $\bar{B}^z_{\text{H}}$ denotes the simple spatial average of $B^z_{\text{H}}(x,\,y)$ over the the Hall sensor surface.
If $B^z_{\text{H}}(x,\,y)$ is uniform and  $S_{\text{H}}(x,\,y)$ is normalized according to $\bar{S}_{\text{H}} = \int_{0}^{L_x} \text{d}x\, \int_{0}^{L_y} \text{d}y\, S_{\text{H}}(x,\,y) = 1$, then \refeqs{}~\refeqnum{Eq::HallVoltage_Uniform} and \refeqnum{Eq::HallVoltage_Nonuniform} are equivalent, as expected. 
In \refeq{}~\refeqnum{Eq::EffectiveMagneticField}, we have neglected any $z$-dependence of the Hall sensor by assuming $L_z$ is very thin.
This approach is valid for the Hall sensor from the reference experiment, which uses a 2D electron gas for the conducting Hall sensor channel \cite{2017_Kumar_ChiralityInducedSpinPolarizationPlacesSymmetry}.

An analytical expression for the Hall voltage arising from a nonuniform magnetic field over a thin ($L_z \rightarrow 0$) Hall sensor with infinite extent in the $x$ dimension has been presented \cite{1994_Mantorov_InterpretationHallEffectMeasurements}.
It was shown that magnetic fields at $|x-x_{\text{probe}}| > 1.5 L_y$ contribute minimally to the Hall voltage, where $[x_{\text{probe}},\, 0]$ and $[x_{\text{probe}},\, L_y]$ are the coordinates of two small Hall sensor probes.
For this reason, the Hall sensor sensitivity is negligible beyond $|x-x_{\text{probe}}| > 1.5 L_y$, so we assume the Hall voltage expression for an infinitely long Hall sensor can apply to the finite Hall sensor from the reference experiment, which had dimensions $L_x = 12.5 L_y \gg 1.5 L_y$.
We still limit the magnetic field to a finite surface given by $x \in [0,\,L_x]$ and $y \in [0,\,L_y]$.

% #NOTE: To attempt to formalize this justification, we have:
% $\int_{\text{infinite}} S_{\text{H,infinite}} B_{\text{finite}} = \int_{\text{finite}} S_{\text{H,infinite}} B_{\text{finite}} \approx \int_{\text{finite}} S_{\text{H,infinite}} B_{\text{infinite}}$, so we assume $\int_{\text{finite}} S_{\text{H,infinite}} B_{\text{finite}} \approx \int_{\text{finite}} S_{\text{H,finite}} B_{\text{finite}}$.
% Then $S_{\text{H,finite}} \approx S_{\text{H,infinite}}$, where $S_{\text{H,finite}}$ is some exact sensitivity for a finite Hall sensor.

\begin{figure}[htbp!]
    \captionsetup[subfloat]{farskip=0pt,captionskip=0pt}
    \subfloat[][]{\includegraphics{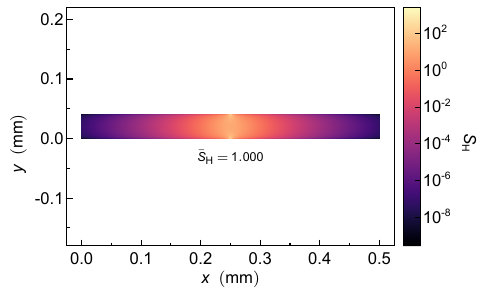}\label{Fig::HallSensor_Sensitivity}}\\
    \subfloat[][]{\includegraphics{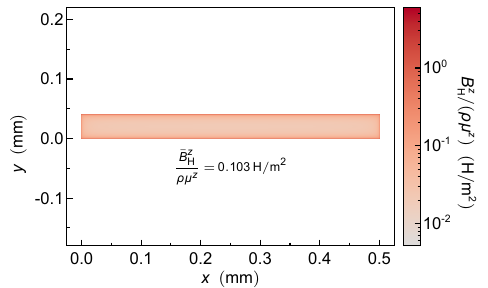}\label{Fig::HallSensor_MagneticField}}\\
    \subfloat[][]{\includegraphics{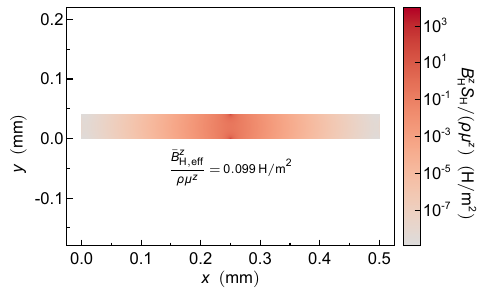}\label{Fig::HallSensor_WeightedMagneticField}}
    \caption{
        (a) Spatially resolved Hall sensor sensitivity according to \refeq{}~\refeqnum{Eq::HallSensorSensitivity}.
        (b) Spatially resolved magnetic field from a surface density of dipoles according to \refeq{}~\refeqnum{Eq::ZDipoleSurface_Approximation}.
        (c) Spatially resolved magnetic field from (b) when weighted by the sensitivity from (a).
    }
    \label{Fig::HallSensor}
\end{figure}

By comparing \refeq{}~\refeqnum{Eq::HallVoltage_Nonuniform} to the Hall voltage expression that results from applying a magnetic field constrained to $x \in [0,\,L_x]$ and $y \in [0,\,L_y]$ to a Hall sensor with infinite $x$ extent \cite{1994_Mantorov_InterpretationHallEffectMeasurements}, we write the Hall sensor sensitivity as
\begin{equation}\label{Eq::HallSensorSensitivity}
    S_{\text{H}}(x,\,y) = \frac{L_x}{L_y} \, \frac{\cosh{ \frac{\pi (x - x_{\text{probe}})}{L_y}} \sin{ \frac{\pi y }{L_y}}}{\sinh^2{ \frac{\pi (x - x_{\text{probe}})}{L_y}} + \sin^2{ \frac{\pi y }{L_y}}}\eqperiod{}
\end{equation}
Taking $x_{\text{probe}} = L_x/2$ results in $S_{\text{H}}(x,\,y)$ symmetric about $x=L_x/2$ and $y=L_y/2$.
In such a case, plots of $S_{\text{H}}(x,\,y)$, $B^z_{\text{H}}(x,\,y)$, and $S_{\text{H}}(x,\,y) B^z_{\text{H}}(x,\,y)$ are shown in \reffig{}~\ref{Fig::HallSensor}, where $B^z_{\text{rect},z}(x,\,y,\,\Delta z)$ [from \refeq{}~\refeqnum{Eq::ZDipoleSurface_Approximation}] was used for $B^z_{\text{H}}(x,\,y)$.
% #NOTE: The resulting $\bar{B}_{\text{H}}$ and $\bar{B}_{\text{H,eff}}$ differ by less than $5\%$.

\begin{figure}[htbp!]
    \captionsetup[subfloat]{farskip=0pt,captionskip=0pt}
    \subfloat[][]{\includegraphics{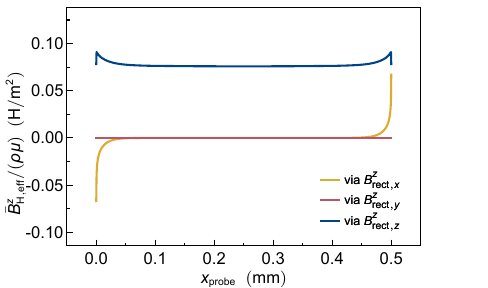}\label{Fig::HallSensor_ProbeInfluence}}\\
    \subfloat[][]{\includegraphics{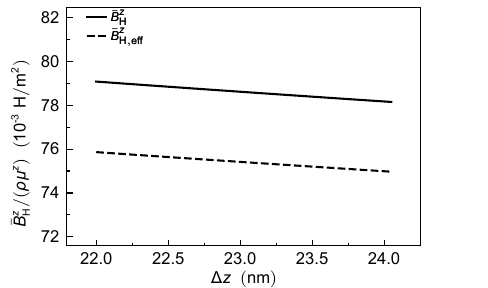}\label{Fig::HallSensor_Comparison}}
    \caption{
        (a) Weighted spatial average magnetic field returned by a Hall sensor as a function of the Hall sensor probe locations, where the magnetic field is from a surface density of magnetic dipoles with magnetic moment $\mu^x \hat{\bm{x}}$ (yellow), $\mu^y \hat{\bm{y}}$ (red), and $\mu^z \hat{\bm{z}}$ (blue), with $\mu^x = \mu^y = \mu^z = \mu$.
        (b) Comparison between the weighted spatial average of a magnetic field returned by a Hall sensor (dashed line) and the simple spatial average of a magnetic field (solid line), where the magnetic field is from a surface density of magnetic dipoles with magnetic moment $\mu^z \hat{\bm{z}}$ given by \refeq{}~\refeqnum{Eq::ZDipoleSurface_Approximation}.
    }
    \label{Fig::HallSensor_Sweep}
\end{figure}

The reference experiment has two sets of Hall sensor probes that are positioned according to $x_{\text{probe}}\approx 0.5 L_x \pm 0.2L_x$ \cite{2017_Kumar_ChiralityInducedSpinPolarizationPlacesSymmetry}, which breaks the symmetry of $S_{\text{H}}(x,\,y)$ about $x=L_x/2$. \Reffig{}~\ref{Fig::HallSensor_ProbeInfluence} shows that the contributions to $\bar{B}^z_{\text{H,eff}}$ from $B^z_{\text{rect},x}(x,\,y,\,\Delta z)$, $B^z_{\text{rect},y}(x,\,y,\,\Delta z)$, and $B^z_{\text{rect},z}(x,\,y,\,\Delta z)$ when $x_{\text{probe}}$ ranges from $0$ to $L_x$. Because of symmetries of the Hall sensor sensitivity and the magnetic fields, the contribution from $B^z_{\text{rect},y}(x,\,y,\,\Delta z)$ is zero regardless of $x_{\text{probe}}$, but the contribution from $B^z_{\text{rect},x}(x,\,y,\,\Delta z)$ can be nonzero.
However, when $x_{\text{probe}}\approx 0.5 L_x \pm 0.2L_x$, the contribution from $B^z_{\text{rect},x}(x,\,y,\,\Delta z)$ is negligible compared to that of $B^z_{\text{rect},z}(x,\,y,\,\Delta z)$, provided $\mu^x \approx \mu^z$.
Thus, similar to Section~\ref{Sec::DipoleSurfaceMagneticField}, we neglect $B^z_{\text{rect},x}(x,\,y,\,\Delta z)$ and $B^z_{\text{rect},y}(x,\,y,\,\Delta z)$.

% #NOTE: A Hall sensor with $x_{\text{probe}} = L_x/2$ has a sensitivity that is symmetric about $x=L_x/2$ and $y=L_y/2$.
% Because $B^z_{\text{rect},x}(x,\,y,\,\Delta z)$ is antisymmetric about $x=L_x/2$ and $B^z_{\text{rect},y}(x,\,y,\,\Delta z)$ is antisymmetric about $y=L_y/2$, the corresponding $\bar{B}^z_{\text{H,eff}}$ value for each case is zero by symmetry. 
% Antisymmetric probes relative to the magnetic-field extents, i.e., $x_{\text{probe}} \neq L_x/2$, leaves the symmetry of $S_{\text{H}}(x,\,y)$ about $y=L_y/2$ intact, but breaks the symmetry of $S_{\text{H}}(x,\,y)$ about $x=L_x/2$.
% This means $B_{\text{rect},y}(x,\,y,\,\Delta z)$ still leads to
% $\bar{B}_{\text{H,eff}} = 0$, but $B_{\text{rect},x}(x,\,y,\,\Delta z)$ now leads to $\bar{B}_{\text{H,eff}} \neq 0$. 

% #NOTE: When approximating a finite Hall sensor as one with an infinite $x$ extent and a finite magnetic-field extent, it is most reasonable to use \refeq{}~\refeqnum{Eq::HallSensorSensitivity} when $1.5L_y<x_{\text{probe}}<(L_x-1.5L_y)$.
% Otherwise, the finite magnetic field would no longer extend to the point of negligible sensor sensitivity, and one might expect the sensitivity to be altered by the proximity of the probes to the sensor boundaries.
% Nonetheless, we include $0<x_{\text{probe}}<L_x$ in \reffig{}~\ref{Fig::HallSensor_ProbeInfluence} to illustrate the symmetries discussed, which are otherwise less apparent.

\Reffig{}~\ref{Fig::HallSensor_Comparison} shows $\bar{B}^z_{\text{H,eff}}$ and $\bar{B}^z_{\text{H}}$ for various $\Delta z$, where $B^z_{\text{rect},z}(x,\,y,\,\Delta z)$ [from \refeq{}~\refeqnum{Eq::ZDipoleSurface_Approximation}] was used for $B^z_{\text{H}}(x,\,y)$ and dimensions were selected to correspond to the reference experiment.
The maximum difference between $\bar{B}^z_{\text{H,eff}}$ and $\bar{B}^z_{\text{H}}$ at the relevant dimensions is $4.15\%$.
Ultimately, the effect of nonuniformity on the Hall voltage is inconsequential for the magnetic field relevant to the reference experiment such that $\bar{B}^z_{\text{H,eff}} \approx \bar{B}^z_{\text{H}}$, which amounts to the assumption used in the main text.

% #NOTE: In \reffig{}~\ref{Fig::HallSensor_Comparison}, starting from $B^z_{\text{ribbon},z}(x,\,y,\,\Delta z)$ and $B^z_{\text{cross},z}(x,\,y,\,\Delta z)$ we compare $\bar{B}^z_{\text{H}}$ and $\bar{B}^z_{\text{H,eff}}$.
% The lower bounds of $\bar{B}^z_{\text{H}}$ and $\bar{B}^z_{\text{H,eff}}$ coincide because $B^z_{\text{ribbon},z}(x,\,y,\,\Delta z)$ is only dependent on $y$ and the Hall voltage is not significantly impacted by $y$ nonuniformity when $L_x \gg 1.5L_y$ \cite{1994_Mantorov_InterpretationHallEffectMeasurements}.
% (More precisely, the Hall voltage of a sensor with infinite extent is not impacted by $y$ nonuniformity, while $B^z_{\text{cross},z}(x,\,y,\,\Delta z)$ is dependent on $x$ and $y$.)
% However, the separation between the Hall sensor probes and the $x$ nonuniformity of $B^z_{\text{cross},z}(x,\,y,\,\Delta z)$ reduces the impact of these edge effects on $\bar{B}^z_{\text{H,eff}}$ compared to $\bar{B}^z_{\text{H}}$.

\section{Quasistatic approximation for calculation of magnetic fields}

Here, we show that the intermediate- and far-zone contributions to a magnetic field from a time-varying magnetic dipole moment are not essential for our application compared to the near-zone contribution used in the main text.

A dipole at position $\bm{r}' = [x',\; y',\; z']$ with a $\hat{\bm{z}}$-directed magnetic moment $\mu_m^z(t)$ generates a magnetic field at $\bm{r} = [x,\,y,\,z]$ with a $\hat{\bm{z}}$ component given by \cite{1998_Heras_ExplicitExpressionsElectric}
\begin{equation}\label{Eq::DipoleMagneticField_FullContribution}
    \begin{split}
        B_m^z(\bm{r},\,\bm{r}',\,t) ={}& \frac{\mu_0}{4\pi} \mu_m^z(t') \bigg[ \frac{3 (z-z')^2}{|\bm{r}-\bm{r}'|^5} - \frac{1}{|\bm{r}-\bm{r}'|^3}\bigg] \\
        &+ \frac{\mu_0}{4\pi c}\, \frac{\text{d}}{\text{d}t'} \mu_m^z(t') \bigg[ \frac{3 (z-z')^2}{|\bm{r}-\bm{r}'|^4} - \frac{1}{|\bm{r}-\bm{r}'|^2}\bigg] \\
        &- \frac{\mu_0}{4\pi c^2}\, \frac{\text{d}^2}{\text{d}(t')^2} \, \mu_m^z(t') \frac{(x-x')^2 + (y-y')^2}{|\bm{r}-\bm{r}'|^3}\eqcomma{}
    \end{split}
\end{equation}
where $t' = t - |\bm{r} - \bm{r}'|/c$ is the propagation-delayed time and $c$ is the speed of light. Like in the main text, we neglect any time delays from propagation by assuming $t' = t$.
The first, second, and third term of \refeq{}~\refeqnum{Eq::DipoleMagneticField_FullContribution} correspond to the near-, intermediate-, and far-zone contributions, respectively.
% #NOTE: We only consider the magnetic field generated by a $\hat{\bm{z}}$-directed magnetic moment because near- and intermediate-field contributions from the $\hat{\bm{x}}$ and $\hat{\bm{y}}$ components of dipoles can be shown to be zero using a similar analysis to that from Section~\ref{Sec::DipoleSurfaceMagneticField}.

\begin{figure}[htbp!]
    \captionsetup[subfloat]{farskip=0pt,captionskip=0pt}
    \subfloat[][]{\includegraphics{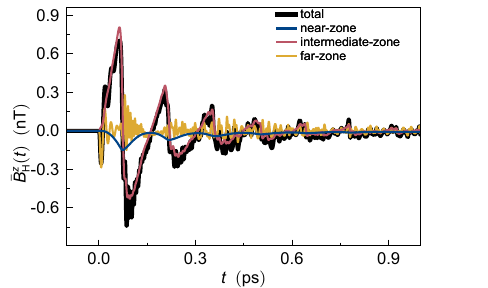}\label{Fig::FarFieldMagneticField_NoFilter}}\\
    \subfloat[][]{\includegraphics{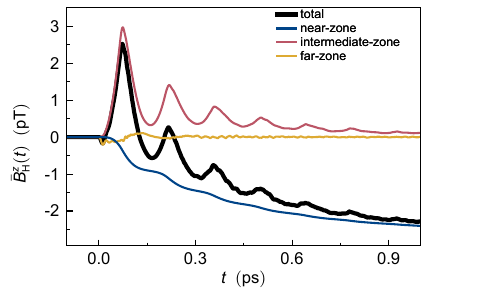}\label{Fig::FarFieldMagneticField_LowPass}}
    \caption{
        Near-, intermediate-, and far-zone contributions to the magnetic field from the spin polarization calculated for our monolayer model using parameters from the main text, when an $RC$ low-pass filter is (a) absent and (b) present.
    }
    \label{Fig::FarFieldMagneticField}
\end{figure}

First, we compute the spin polarization of a right-chiral molecule with one lead using parameters given by the main text.
Next, we apply this spin polarization to the monolayer model described in the main text and calculate the magnetic field of the resulting magnetic dipole distribution using \refeqs{}~\refeqnum{Eq::DipoleMagneticField_FullContribution} and \refeqnum{Eq::GeneralDipoleSurface_Average}.
The result is shown in \reffig{}~\ref{Fig::FarFieldMagneticField_NoFilter}, which appears to indicate that the intermediate-zone contribution has the largest impact on the magnetic field.
However, the presence of a large capacitance in the reference experiment \cite{2017_Kumar_ChiralityInducedSpinPolarizationPlacesSymmetry} dictates the timescales of the system according to time constant $\tau=RC$, which amounts to an $RC$ low-pass filter. 
Hence, the calculated magnetic-field signatures that correspond to the experimental measurement are better represented as a low-passed equivalent.

An $RC$ low-pass filter with $\tau = 10\;\text{ps}$ largely removes the contribution from the intermediate zone because of its oscillating sign, as shown by \reffig{}~\ref{Fig::FarFieldMagneticField_LowPass}.
Thus, the near-zone contribution sustains the largest magnetic-field magnitude.
Ultimately, although the intermediate- and far-zone contributions may influence the Hall signal, they are not expected to impact the low-passed peak magnetic fields by an order of magnitude, and are neglected for the purposes of our qualitative comparison.

\section{Approximation of solenoid as a point dipole}

\begin{figure}
    \captionsetup[subfloat]{farskip=0pt,captionskip=0pt}
    \subfloat[][]{\includegraphics{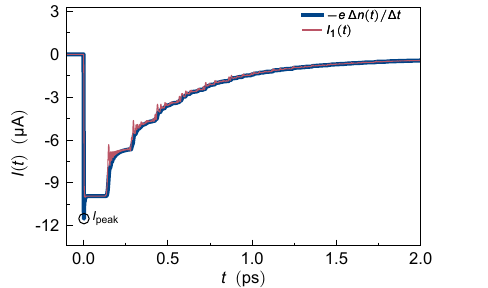}\label{Fig::SolenoidCurrent_MaximumCurrentEstimate}}\\
    \subfloat[][]{\includegraphics{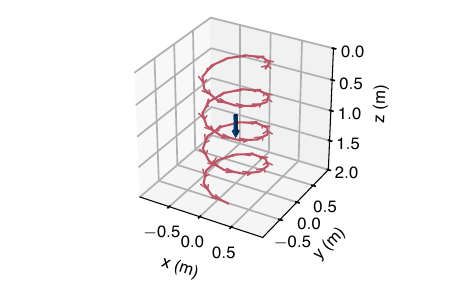}\label{Fig::SolenoidCurrent_Diagram}}
    \caption{
        (a) The aggregate change in charge per unit time throughout the molecule (blue) and the current between sites 1 and 2 (red) for a chiral molecule with parameters given by the main text.
        (b) Diagram of a solenoid (red) with current $I_{\text{peak}}$ and the corresponding approximation consisting of a magnetic dipole (blue) with magnetic moment $\mu^z_I = s M \pi r^2 I_{\text{peak}}$.
    }
\end{figure}

First, we show that the aggregate change in charge per unit time closely approximates the current entering the molecule.
The aggregate change in charge per unit time, defined as $-e\,\Delta n(t) / \Delta t$ by the main text, was calculated using the electron density.
The current between sites 1 and 2 was calculated in the same manner as Section~\ref{Sec::SymmetryCurrent} and denoted $I_1(t)$.
Both of these quantities are shown in \reffig{}~\ref{Fig::SolenoidCurrent_MaximumCurrentEstimate}, which illustrates a high level of agreement.
The temporal peak of $-e\,\Delta n(t) / \Delta t$, denoted $I_{\text{peak}}$, differs from the temporal peak of $I_1(t)$ because the former includes the charge added to site 1, whereas the latter only captures the current between sites 1 and 2.
\\[0ex]
\indent Next, we consider the magnetic flux over the $L_x=500\;\mathrm{\upmu m}$ by $L_y = 40\;\mathrm{\upmu m}$ sensor surface defined by the main text, as generated by a solenoid with current $I_{\text{peak}}$ and by an approximation of this solenoid in the form of a point dipole with a magnetic moment $\mu^z_I = s M \pi r^2 I_{\text{peak}}$.
The solenoid has dimensions equal to the chiral-molecule dimensions from the main text, and the dipole is located halfway along the chiral molecule, as shown in \reffig{}~\ref{Fig::SolenoidCurrent_Diagram}.
The dipole and solenoid are centered in the $xy$ plane with respect to the sensor surface, but have a $z$ offset from the sensor surface.
Considering each atom within the chiral molecule as a vertex on the solenoid, we use the Magpylib Python library \cite{2020_Ortner_MagpylibFreePythonPackage} to calculate the normalized fluxes from the solenoid and dipole as $-1.44\times10^{-9} \; \text{T/A}$ and $-1.56\times10^{-9} \; \text{T/A}$, respectively, where we have normalized these fluxes by $L_x L_y I_{\text{peak}}$.
These values have an $8.52\%$ difference, so approximating the solenoid as a point dipole is reasonable for our purposes.

\bibliography{ref.bib}